\documentclass[acmsmall]{acmart}
\usepackage{adjustbox}
\usepackage{pdflscape}
\PassOptionsToPackage{table,xcdraw}{xcolor}
\usepackage{colortbl}
\usepackage{multirow}
\usepackage{subfig} 
\usepackage{hyperref}
\usepackage{lscape}
\usepackage{longtable}
\usepackage{graphicx} 
\usepackage{array}

\usepackage{xcolor}

\AtBeginDocument{%
  }

\setcopyright{acmlicensed}
\copyrightyear{2026}
\acmYear{2026}
\acmConference[CSCW '26]{The 2026 ACM Conference on Computer-Supported Cooperative Work and Social Computing}{October 10--14, 2026}{}

\begin{document}

\title[Understanding the Gap Between Stated and Revealed Preferences in News Curation]{Understanding the Gap Between Stated and Revealed Preferences in News Curation: A Study of Young Adult Social Media Users}

\author{Do Won Kim}
\affiliation{
  \institution{University of Maryland}
  \city{College Park}
  \country{United States}}
\email{dowonkim@umd.edu}

\author{Cody Buntain}
\affiliation{
  \institution{University of Maryland}
  \city{College Park}
  \country{United States}}
\email{cbuntain@umd.edu}

\author{Giovanni Luca Ciampaglia}
\affiliation{
  \institution{University of Maryland}
  \city{College Park}
  \country{United States}}
\email{gciampag@umd.edu}

\begin{abstract}
Social media feed algorithms infer user preferences from their past behaviors. Yet what drives engagement often diverges from what users value. We examine this gap between stated preferences (what users say they prefer) and revealed preferences (what their behavior suggests they prefer) among young adults, a group deeply embedded in algorithmically mediated environments. Using a mixed-methods approach combining surveys and interviews with feed curation activities, we investigate: what gaps exist between stated and revealed preferences; how users make sense of these gaps; what values users believe should guide algorithmic curation; and how systems might reflect those values. Participants often found themselves engaging with low-quality content they did not endorse, despite wanting high-quality information. When asked to curate an ideal social media news feed for a hypothetical persona, participants created feeds they considered more satisfying and higher in quality by prioritizing values such as accuracy and diversity. In doing so, they navigated trade-offs between different values, factoring in social relationships and context surrounding the persona. These findings suggest that feed curation is a socially situated process of judging what should be visible and appropriate in shared information spaces. Based on these insights, we offer design directions for bridging the gap between stated and revealed preferences.
\end{abstract}




\maketitle

\section{\MakeUppercase{Introduction}}
Social media news feed algorithms infer user preferences from engagement behaviors, even though users often engage with content they do not actually endorse \cite{AganEtAl2023, Kleinberg2024, Kleinberg2023, MilliEtAl2023a, RathjeEtAl2024a, StewartEtAl2024}. This is referred to as the gap between stated preferences (what users say they want and value) and revealed preferences (what their behaviors suggest they want and value). 

As recommendation algorithms are intermediaries that allocate attention and shape what gains visibility and legitimacy \cite{LazarEtAl2024}, problems can arise when that attention is systematically allocated to content users do not actually value. Users often engage with attention-grabbing content that is provocative or divisive \cite{MilliEtAl2023a, RathjeEtAl2024a}. Algorithms interpret such engagement as endorsement and amplify similar content, potentially distorting the information environment over time \cite{LazerEtAl2024}.

We focus on U.S.-based young adult social media users (i.e., aged 18--24), a group deeply embedded in algorithmically mediated environments \cite{BraghieriEtAl2022, BursztynEtAl2023}. We start by asking two questions: (\textbf{RQ1}) What gaps exist between revealed and stated preferences among these users? (\textbf{RQ2}) How do they make sense of these gaps? 
If such gaps exist, they challenge the assumption that engagement reflects what users truly value, raising questions about which values should guide algorithmic curation and how. Thus, we further ask: (\textbf{RQ3}) How do users identify and navigate different values that shape ideal news feeds? And (\textbf{RQ4}) how can we design news feed recommendation systems to be better aligned with what users explicitly value?

We answer these questions using surveys and interviews with feed curation tasks. To reduce the potential for social desirability bias, participants who showed a gap between stated and revealed preferences were asked to act as recommendation system designers, curating a persona's feed based on stated values. As such, this study frames recommender systems as infrastructures that mediate value-laden decisions, and it offers insights into how these values can be embedded within them.

Although participants stated that they wanted high-quality content, they often engaged with low-quality posts, where quality is operationalized using source reliability ratings by NewsGuard. They attributed the gap to a misalignment with the rankings provided by the news feed algorithm: while users prioritized quality, algorithms prioritized engagement. When asked to curate feeds that align with the values they attribute to a chosen persona, participants created feeds that were both higher in quality and more satisfying than a reference engagement-maximizing feed.  
In curating the feeds, participants were guided by some recurring values. While they prioritized social values such as trustworthiness and diversity, they also balanced these against personally engaging and relevant content tailored to the persona. Social context cues of the persona also influenced curation decisions, suggesting that judgments related to feed curation are socially situated, as users negotiate what should be visible and appropriate in a shared information environment. Finally, participants proposed design ideas for aligning social media feeds with social values. 

This study makes two key contributions. First, this study contributes to research on value-sensitive and user-centered design. Prior work has shown that engagement does not fully represent what users truly want, while platforms nonetheless continue to optimize for engagement for profit-oriented incentives \cite{StewartEtAl2024, Kleinberg2023, RathjeEtAl2024a, MilliEtAl2023a, AganEtAl2023}. Our study demonstrates how stated values of users can be elicited and translated into actionable algorithmic criteria in a dynamic, content-rich environment. Through this process, we identify concrete design pathways for incorporating value-oriented preference elicitation into recommender systems, and outline key implementation opportunities as well as challenges for future work.

Second, we extend research on user participation in algorithmic systems by showing how users perceive, negotiate, and respond to this gap within algorithmically mediated environments \cite{LazerEtAl2024, ParkEtAl2022, MokEtAl2023, ShenEtAl2024, StrayEtAl2023, Gabriel2020, LeeEtAl2019}. Unlike prior work, which focused on UX interventions, rule-based controls, and simple sorting mechanisms \cite{HarambamEtAl, Gobo, LukoffEtAl}, in our study, participants reflected on trade-offs among different values, considering which non-engagement dimensions (such as trustworthiness, diversity, and safety) should be prioritized and how to balance them with the engagement qualities of content (such as entertainment, popularity, and relevance). These findings show that interactions with recommender systems involve everyday negotiations over what counts as socially and individually desirable, and illustrate how systems can better support reflective, value-based interventions in shaping what algorithmic feeds promote.

\section{\MakeUppercase{Related Work}}
We build on two areas of related work: (1) research on the gap between stated and revealed preferences in social media contexts, and (2) HCI and CSCW work on value-aligned algorithms.

\subsection{Gap Between Stated and Revealed Preferences}\label{gap}
According to the dual process theory of decision-making, human decision-making is governed by two distinct information processing systems: System 1, which is fast, automatic, and intuitive, and System 2, which is slower, effortful, and reflective \cite{Kahneman2011}. When System 1 dominates, individuals are more susceptible to cognitive biases and are more likely to act in ways that diverge from their stated preferences \cite{ThalerEtAl2008, AganEtAl2023, Kleinberg2023}. 

This tendency becomes especially pronounced in social media environments. Social media algorithms rely on behavioral data, assuming that what users do accurately reflects what they want \cite{StrayEtAl2023, MorewedgeEtAl2023}. However, engagement-based algorithms are designed to trigger System 1 processing by prioritizing emotionally charged or divisive content that captures immediate attention \cite{RathjeEtAl2024a, RobertsonEtAl2023}. Such designs exploit attention of users and are further reinforced by their limited self-control and habitual engagement \cite{AllcottEtAl2022}. 

Previous research has extensively documented this gap between stated and revealed preferences in social media contexts \cite{StewartEtAl2024, Kleinberg2023, RathjeEtAl2024a, MilliEtAl2023a, AganEtAl2023}. These studies show that users often engage with content they do not consciously value, such as misinformation \cite{StewartEtAl2024}, political divisiveness \cite{RathjeEtAl2024a}, or high emotional valence \cite{MilliEtAl2023a}. This is likely due to a number of human cognitive biases, like inattention \cite{StewartEtAl2024}, automatic behavior \cite{Kleinberg2023}, or time pressures, which can result in choices that encode and reproduce ingroup biases \cite{AganEtAl2023}. Recommender systems are trained on this data, which could further promote this class of content. 

Some CSCW and HCI studies have also pointed to this gap. Lee and Baykal \cite{LeeBaykal2017} argue that algorithms assume users as self-interested agents, overlooking altruistic or socially motivated behaviors. Pyle et al. \cite{PyleEtAl2024} show that even technical experts find it difficult to resist or control the influence of algorithms, suggesting a disconnect between user intent and algorithmic behavior.

Despite substantial evidence documenting gaps between stated and revealed preferences on social media, far less is known about how these gaps manifest among users aged 18–-24, one of the most intensive users of social media. And although social media is the primary source of news in the U.S. and worldwide \cite{st.aubin2024social}, young people often enter civic life with weak or undecided partisan identities, limited engagement with news, and lower levels of political knowledge \cite{VelezEtAl2025}. Consequently, their exposure to political content is shaped less by stable identities or established habits and more by algorithmic curation, increasing their reliance on platform signals when navigating news environments. In this context, engagement-maximizing algorithms may exert a disproportionate influence on what young adults see and consume. Furthermore, even less understood is how these young adults perceive, interpret, and reason about these gaps---insights that are critical for designing recommendation systems that align with underlying values of users rather than transient engagement behaviors. 

In this light, we ask: (\hypertarget{rq1}{\textbf{RQ1}}) What gaps exist between revealed and stated preferences among young adult social media users? And (\hypertarget{rq2}{\textbf{RQ2}}) how do they make sense of these gaps?

\subsection{Value-Aligned Algorithms}  

\subsubsection{Trade-Offs Between Values}
As people hold divergent views on what values algorithmic systems should prioritize \cite{JakeschAtEl2022, NguyenEtAl2024}, designing value-aligned algorithms is challenging. The question is which values are being encoded, and how value conflicts can be addressed in system design.

Research has identified a set of recurring and sometimes conflicting value as alignment goals for recommender systems \cite{StrayEtAl2023, ShenEtAl2024, Zhi-XuanEtAl2024}. Central to this discussion is the tension between algorithms that optimize for revealed preferences and those that optimize for stated preferences. Algorithms based on revealed preferences prioritize delivering relevant content that is immediately rewarding to users \cite{MorewedgeEtAl2023}. Platforms favor these algorithms since they quickly convert user behavior into utility metrics and recommend content without requiring explicit user input and editorial interventions \cite{StrayEtAl2023, Zhi-XuanEtAl2024}. While these algorithms help reduce information overload \cite{AganEtAl2023, Swart2021}, they make automated decisions about what information to deliver \cite{Swart2021}, often compromising other values like diversity and accuracy.

On the other hand, aligning algorithms with the stated preferences of users emphasizes user control and agency \cite{FengEtAl2024}. In this context, stated preferences serve as proxies for what users would choose after careful consideration \cite{Zhi-XuanEtAl2024}. This interpretation of stated preferences is consistent with Gabriel's \cite{Gabriel2020} definitions of desires (what one would desire if rational and informed), interests (what is best for oneself), and values (what is morally right). This is also consistent with Morewedge et al.'s \cite{MorewedgeEtAl2023} concept of `shoulds,' choices that prioritize long-term benefits over immediate rewards. 

In summary, the key trade-off lies between optimizing for immediate engagement and delivering content that aligns with the conscious values of users. While relevance and usefulness are key for delivering timely and engaging content, overemphasis on them can undermine user agency and control. In contrast, algorithms that prioritize the stated preferences of users may enhance user agency and provide content that is more aligned with their values, but they may require more explicit input from users and may not maximize immediate rewards. Thus, finding the right balance between user control and engagement is critical for developing algorithms that enhance user autonomy while still delivering relevant content effectively. 

\subsubsection{Proposed solutions}
Researchers have proposed shifting the balance towards prioritizing user control or agency. For instance, Feng et al. \cite{FengEtAl2024} advocate for providing users with more controls over their algorithmic experiences, such as feed customization and content opt-out options. Lazer et al. \cite{LazerEtAl2024} argue that, in the absence of such controls, users may be more likely to engage with eye-catching content that does not align with their stated preferences.

Along these lines, a growing body of work has tested various interface-level mechanisms for shaping algorithmic feeds through user controls \cite{lukoff2023switchtube, LukoffEtAl, HarambamEtAl, Gobo, lyngs2020just, EkstrandEtAl2015}. For example, Ekstrand et al. \cite{EkstrandEtAl2015} allowed users to select among different recommendation algorithms on MovieLens, showing how users engaged with this feature to experiment with alternatives before settling on their preferred algorithm. Lukoff et al. \cite{LukoffEtAl} examine how UX features such as autoplay and endless recommendations influence users’ sense of agency on platforms like YouTube, and propose design strategies to support user control. Harambam et al. \cite{HarambamEtAl} similarly introduce feed controllers that allow users to adjust topic-level preferences using sliders and switches, enabling users to influence the composition of their news feeds. Another example is Gobo \cite{Gobo}, a system that allows users to apply item-level, rule-based filters (e.g., seriousness, politics, or rudeness) to steer recommendation outcomes across multiple social media feeds.

However, there are limitations to control-based approaches. Users often struggle to effectively use control features and may have a limited understanding of how algorithmic systems function and how their actions shape feed outcomes \cite{JhaverEtAl2023, BradyEtAl2023b, JunejaEtAl2023, StormsEtAl2022, konig_challenges_2024}. These challenges suggest that interface-level controls alone may be insufficient for aligning feeds with the underlying values of users.

Another approach is to include user preferences as a goal of algorithm optimization \cite{MorewedgeEtAl2023, StrayEtAl2023}. Stray et al. \cite{StrayEtAl2023} argue for using survey data to train algorithms to better reflect user values but also note that existing methods typically rely on an item-by-item approach, asking users for their stated preferences on a small number of individual content items, which is inherently limited. It is infeasible to ask users about every single item they encounter, it places excessive cognitive burden on them, and it is difficult to generalize beyond the specific content rated \cite{StrayEtAl2023}. Instead, there is a need for mechanisms to effectively capture the broader stated preferences of users. 

Prior work has begun to explore this direction. Feng et al.~\cite{FengEtAl2024} elicit user preferences in an open-ended manner, allowing participants to freely express what they want from algorithmic systems. While this approach surfaces a wide range of user input, these inputs are largely discussed as generic “preferences,” without closely examining what kinds of values users are invoking, how they understand those values, or how different preferences relate to one another. In contrast to an open-ended approach, Rathje et al.~\cite{RathjeEtAl2024a} use a fixed set of predetermined categories, such as divisiveness, hatefulness, and accuracy. This enables systematic comparison across users and content, but constrains participants into a fixed value taxonomy, which may not capture how users would describe themselves what matters to them or reason about their feeds in their own terms.

Therefore, research gaps in this area are threefold. First, while prior work has taken important steps toward eliciting user preferences, existing approaches are limited in terms of either their generalizability or their exhaustiveness. Second, even when relevant value dimensions are identified at the proper level of granularity, little is known about how users weigh them against one another. Third, while a growing body of work has proposed design mechanisms for increasing user control in recommender systems, there remains substantial room to further explore how the value priorities articulated by users can be translated into concrete ranking mechanisms. In summary, many open questions remain about what broader values preferences entail and how they can be incorporated into algorithmic design. 

To fill these gaps, this study asks: (\hypertarget{rq3}{\textbf{RQ3}}) How do participants identify and navigate different values that shape ideal news feeds? (\hypertarget{rq4}{\textbf{RQ4}}) And how can recommendation systems be designed to better align with the stated preferences and values of users? 

\section{\MakeUppercase{methods}}
\begin{figure} 
    \centering
    \includegraphics[width=0.95\textwidth]{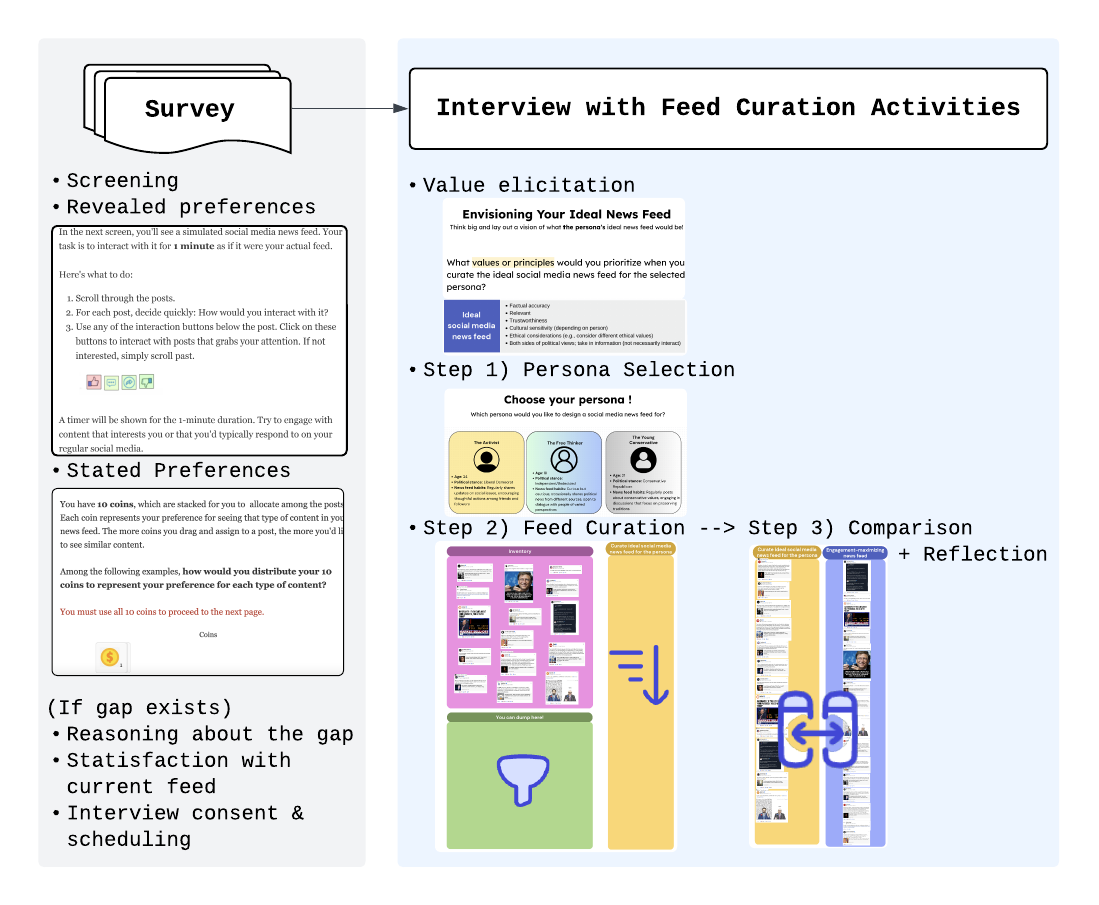}
    \caption{Study overview: participants completed a survey, followed by an interview with feed curation activities.} 
    \Description{Study overview: participants first completed a survey, followed by an interview with feed curation activities.} 
    \label{fig:method_flow}
\end{figure}

This study combined a survey with interviews that incorporated interactive feed curation tasks, as shown in Fig.~\ref{fig:method_flow}, which outlines the overall study flow. Detailed procedures are described in the following subsections. The study was reviewed and approved by the Institutional Review Board (\#2229611-2). 

\subsection{Data}
\subsubsection{Timeline}
Data collection consisted of two rounds, each involving surveys followed by interviews within approximately 7 days. The first-round survey was conducted from October 14 to 18, 2024, with interviews scheduled individually between October 15 to 25. Since the initial round resulted in 10 females and only 2 males, a second round of recruitment was conducted to enhance sample balance. The second-round survey took place from December 14 to 18, 2024, with follow-up interviews held between December 15 and 21.

\subsubsection{Sample Recruitment}
A pool of active social media users aged 18--24 was recruited via Prolific, a research participant recruiting platform. Using pre-screener options provided by Prolific, we targeted participants based on the following criteria: Age (18--24), Nationality (United States), Weekly device usage (More than 2--6 times a week), Social media (Facebook, YouTube, Twitter, Linkedin, Instagram, Reddit, Snapchat, TikTok), Record Video (Yes), Record Audio (Yes), Weekly working hours on Prolific (More than 6 hours). In the second round of data collection, we added an additional pre-screener criterion (Male).

In the survey, participants who failed the attention check or did not use social media for news at least a few times a week were screened out and compensated \$1.50 USD. Those who passed the screening and completed the survey received \$3.00 USD.

\subsubsection{Sample Attrition}
Table \ref{table:table4} shows the sample attrition. Among the 48 participants who initiated the survey, 14 were ineligible and screened out. Participants who did not exhibit gaps between stated and revealed preferences ($n=4$) were intentionally excluded, as our aim was not to demonstrate the existence of such gaps, a question already well established in prior research (see Section~\ref{gap}). Instead, our focus was on less explored questions: where these gaps manifest among young adult social media users, how they understand the gap in relation to algorithmic curation, and how they articulate their value priorities. The interview was thus designed to probe these questions with participants for whom the gap was evident.

\begin{table} 
\caption{Sample attrition}
\centering
\begin{tabular}{|l|c|}
\hline
\rowcolor[HTML]{F3F3F3} \textbf{Stage} & \textbf{N} \\ \hline
Survey started & 48 \\ \hline
\emph{\hspace{2em}No consent or failed attention check} & 7 \\ \hline
\emph{\hspace{2em}Screened out (not using social media for news (n=3) or no gap (n=4))} & 7 \\ \hline
\emph{\hspace{2em}Declined interview consent} & 6 \\ \hline
Completed survey and consented to interview & 28 \\ \hline
\emph{\hspace{2em}No-show or did not schedule interview} & 7 \\ \hline
 Completed interview & 21 \\ \hline
\textbf{Final sample used in analysis} & \textbf{20} \\ \hline
\end{tabular}
\label{table:table4}
\Description{Summary of participant attrition from initial survey to final interview sample.}

\end{table}

Of the participants eligible for the study, 6 declined to participate in the post-survey interview, and 7 either failed to schedule a follow-up or did not attend their appointment. Among the 21 who completed the interviews, one was excluded due to factors that compromised data quality. These included internet connectivity disruptions, limited familiarity with the U.S. context, and difficulties comprehending the assigned tasks. Thus, the final sample consisted of 20 participants, and all analyses reported in this paper are based on data from these 20 participants. 

\begin{table} 
\caption{Demographics of Final Sample}
\resizebox{0.45\textwidth}{!}{
\begin{tabular}{|c|c|c|c|}
\hline
\rowcolor[HTML]{F3F3F3} 
\textbf{Variable}                       & \textbf{Value}     & \textbf{N} & \textbf{\%} \\ \hline
                                        & 19                 & 1          & 5        \\ \cline{2-4} 
                                        & 20                 & 4          & 20        \\ \cline{2-4} 
                                        & 21                 & 2          & 10        \\ \cline{2-4} 
                                        & 22                 & 7          & 35          \\ \cline{2-4} 
                                        & 23                 & 3          & 15         \\ \cline{2-4} 
\multirow{-6}{*}{Age}                   & 24                 & 3          & 15        \\ \hline
                                        & Female             & 10         & 50        \\ \cline{2-4} 
\multirow{-2}{*}{Sex}                   & Male               & 10         & 50        \\ \hline
                                        & Asian              & 2          & 10         \\ \cline{2-4} 
                                        & Black              & 7          & 35        \\ \cline{2-4} 
                                        & Other              & 5          & 25         \\ \cline{2-4} 
\multirow{-4}{*}{Ethnicity}             & White              & 6          & 30        \\ \hline
                                        & Yes                & 12         & 60        \\ \cline{2-4} 
                                        & No                 & 7          & 35        \\ \cline{2-4} 
\multirow{-3}{*}{Student status}        & Etc (data expired) & 1          & 5         \\ \hline
                                        & Full-Time          & 6          & 30          \\ \cline{2-4} 
                                        & Part-Time          & 6          & 30          \\ \cline{2-4} 
\multirow{-3}{*}{Employment status}     & Unemployed/Other   & 8          & 40          \\ \hline
                                        & Democrat           & 8          & 40        \\ \cline{2-4} 
                                        & Independent        & 6          & 30        \\ \cline{2-4} 
                                        & Republican         & 4          & 20         \\ \cline{2-4} 
\multirow{-4}{*}{Political affiliation} & Other              & 2          & 10        \\ \hline
                                        
\end{tabular}
}
\label{table:table5}
\Description{Demographics of Final Sample}
\end{table}

\subsubsection{Final Sample}
Table \ref{table:table5} summarizes the demographics of our final sample. The sample was evenly balanced by sex, with 10 females and 10 males. It had a higher proportion of individuals who self-identified as Democrats (40\%) and Independents/Others (40\%). Only 20\% of the participants identified as Republicans. As a point of comparison, we cross-checked our sample distribution using the 2020 American National Election Studies \cite{ANES2020}, which draws on a nationally representative sample of U.S. citizens. Among respondents aged 19 to 24 ($n = 281$), 41\% identified as Democrats, 40\% as Independents/Others, and 19\% as Republicans, based on self-reported partisanship. 

\subsection{Survey}
We designed the survey with two purposes: to identify participants who showed a gap between stated and revealed preferences (RQ1), and to elicit initial mental models of participants vis-a-vis algorithms and their early reflections on the gap (RQ2). The full questionnaire is in the Appendix~\ref{appendix:survey}.

To measure revealed preferences, participants were shown a mock social media feed containing 8 posts across various source types (see Table \ref{tab:table1} for details on content selection of the mock news feed). They had 1 minute to scroll through the feed and interact with the posts, just as they would on social media. (Woods et al. \cite{WoodsEtAl2022} found that users view 3 posts in 20 seconds on Facebook. Thus, for our 8 posts, it rounds up to 60 seconds.) A timer was displayed, and they were instructed to engage with the posts instinctively and naturally. They could scroll past any posts that did not catch their attention. The number of engagements with these posts served as a measure of their revealed preferences. 

To measure stated preferences, we adapted a budget allocation approach from economics that elicits stated preferences by asking respondents to distribute a fixed amount of budgets across competing options to reveal relative valuations \cite{Schlapfer2017}. Applying this approach, participants were again presented with 8 different posts but across the same categories. Participants were instructed to allocate 10 coins among the posts based on how much they would like to see similar content in their ideal social media news feed. They were required to use all 10 coins. The total amount of coins allocated represents their stated preference for that specific content type: the more coins assigned to a post, the more participants want to see similar content in their feed. 

The gap between stated and revealed preferences was measured by identifying cases where participants engaged with low-quality news content (revealed preference) despite allocating only 1 coin or less to that content category (i.e., as a stated preference). This allocation represents 10\% or less of their ideal feed. In other words, rather than capturing all possible forms of misalignment, we focused on cases where participants expressed a desire to avoid a certain content category (low-quality news) but still engaged with it. 

For participants who exhibited gaps as operationalized as such, we asked their reasoning about this gap, their overall satisfaction with their current news feed, and their mental models of how recommendation algorithms work. Finally, we asked about their willingness to participate in a follow-up interview. Those who consented for the interviews then booked their interview slots. 

\subsection{Interview With Feed Curation Activities}
We designed the interviews as participatory design tasks \cite{ParkEtAl2022, JakeschAtEl2022, FengEtAl2024}, asking participants to act as a recommendation algorithm designer, and engage in a design task for a chosen persona based on values they defined for that persona.  
The interview was structured around three steps: (1) Persona selection and value elicitation, (2) feed curation \cite{FengEtAl2024}, and (3) comparison with an engagement-maximizing feed. 

\begin{figure} 
    \includegraphics[height=0.65\textheight]{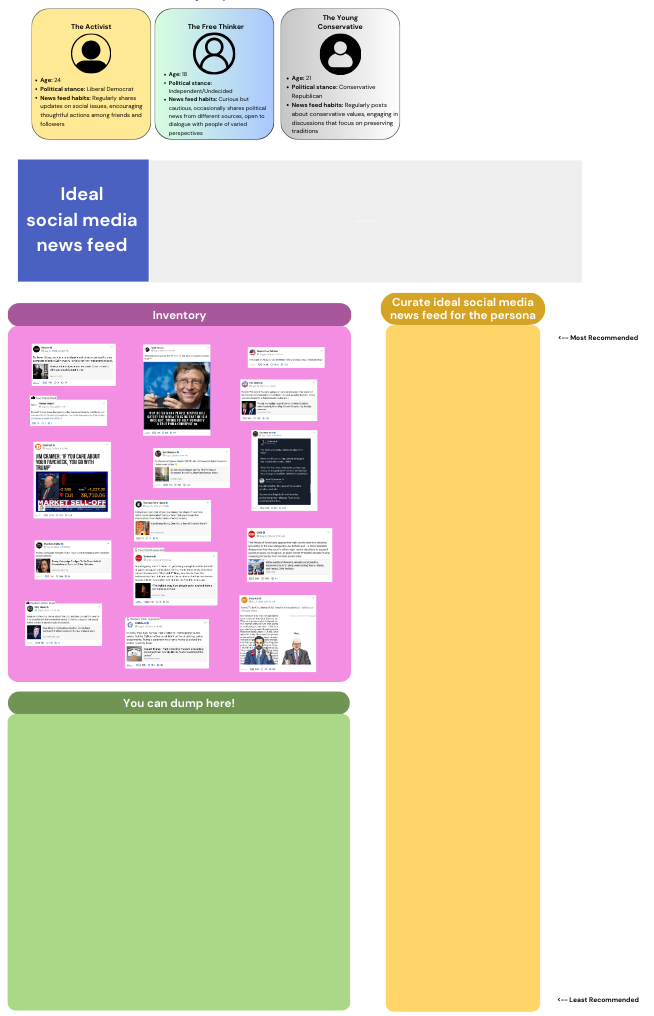}
    \centering
    \Description{Canva board layout: (1) persona profiles at the top, (2) values for an ideal feed (gray box), (3) content inventory (pink box), (4) moderated posts (green box), and (5) curated feed (yellow box).}
    \caption{Canva board layout: (1) persona profiles at the top, (2) values for an ideal feed (gray box), (3) content inventory (pink box), (4) moderated posts (green box), and (5) curated feed (yellow box).}
    \label{fig:Fig1}
\end{figure}

\subsubsection{Step 1: Persona Selection and Value Elicitation.} Participants began the interview by selecting one of three personas for which they would curate a news feed. They were presented with three personas: (1) The Activist (Liberal Democrat, Age 24), who frequently shares updates on social issues and encourages thoughtful discussions among their network, (2) The Free Thinker (Independent/Undecided, Age 18), who is curious but cautious, occasionally shares political news from different sources and is open to dialogue across political perspectives, and (3) The Young Conservative (Conservative Republican, Age 21) who regularly posts about conservative values and engages in discussions focused on preserving traditions. 

Then, participants envisioned the ideal social media space for that persona. They were instructed to think aloud about the values that should guide their ideal news feed curation, and to explain why those values might be important for the selected persona. They were also asked to envision how the chosen persona would engage with content in the ideal social media space. 

The motivation for this step was twofold. First, framing the task from the perspective of a recommendation system designer and asking participants to reason about an ideal news feed for a hypothetical persona was intended to elicit normative judgments about what should guide social media news feed design, rather than expressions of the personal preferences of participants. This approach reduced social desirability bias by avoiding direct evaluation of their own feeds or posting behaviors, while using personas as reflective prompts rather than proxies for participants themselves. Second, presenting multiple personas within the same age range enabled participants to explore how expectations for news feeds might vary across different types of users while holding broader generational context constant, supporting more inclusive design outcomes.

\subsubsection{Step 2: Feed Curation.} 
Participants next transitioned into the role of a recommendation system designer, deciding which items to show in which order in the feed based on the values they outlined in the previous activity. Specifically, they were tasked with curating items from the full inventory (see Section \ref{ContentInventory} for details on its construction) to create an ideal feed for the selected persona. In doing so, they aimed to reflect the vision they had articulated in the earlier phase, shown in the gray box labeled ‘Ideal social media news feed’ in Fig. \ref{fig:Fig1}. 

\begin{figure} 
    \includegraphics[height=0.6\textheight]{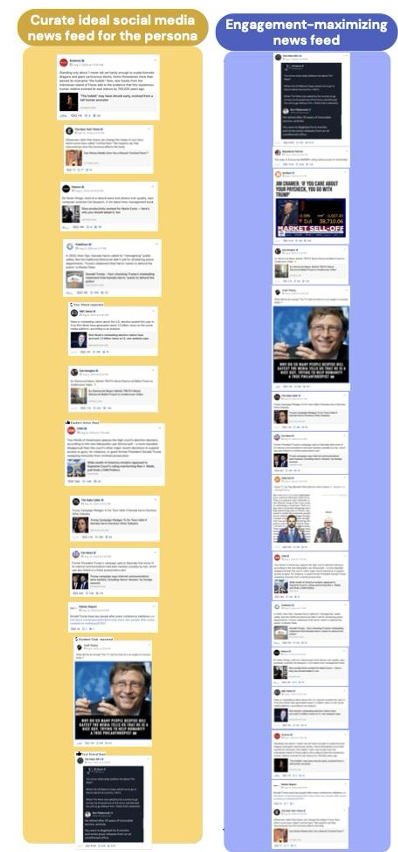}
    \centering
    \Description{An example of a Canva board after a participant has completed the feed curation activity.}
    \caption{An example of a Canva board after a participant has completed the feed curation activity.}
    \label{fig:Fig2}
\end{figure}

Participants were asked to explain not only their ranking decisions but also their reasons for `dumping’ items (as indicated by the green box in Fig. \ref{fig:Fig1}). They were encouraged to elaborate on how their curation decisions aligned with elicited values, as well as to share overall reflections on the activity. Having participants act as a de facto stated preference-based algorithm, moderating and ranking content based on stated values, provides insights into how stated preferences can be translated into actual content ranking. 

\subsubsection{Step 3: Comparing the Curated Feed with an Engagement-Maximizing Feed.} 
After curating the feed, the `Engagement-max Feed’ was shown to participants. This feed served as a simplified baseline that ranks posts by aggregate engagement metrics from the original posts, approximating a feed that prioritizes content likely to elicit broad user interaction. Although this baseline neither personalizes content to the selected persona nor captures the full complexity of modern recommendation algorithms, it provides an interpretable comparison point, which allows participants to see how an engagement-oriented feed may diverge from their value-driven feeds, without the confounding opacity or idiosyncrasies of real-world personalized feeds.

When comparing their curated feeds with the engagement-maximizing feed side-by-side, as shown in Fig. \ref{fig:Fig2}, participants shared their observations on the differences between the two and reflected on how these distinct approaches to content curation might influence users. They then discussed how the exercise influenced their view on news feed algorithms. Participants were also asked to consider what would be needed to bring their ideal feed to life, and any challenges or opportunities in implementing such a feed. 

Each interview lasted approximately one hour on Zoom, and participants were compensated with a \$40 USD bonus payment on Prolific upon completion. The interview transcripts were initially auto-generated by Zoom and later manually refined.
Note that before formal data collection, we conducted three pilot interviews to refine the interview protocol and ensure clarity in eliciting thought processes of participants during feed curation. Pilot data were used exclusively for improving the protocol and were not included in any analyses or results presented in this study. For the full interview protocol, see Appendix \ref{appendix:protocol}.

\subsection{Task Material}\label{ContentInventory}
The task material used for mock feeds in both the survey and interviews was carefully curated to cover a diverse range of dimensions outlined in Table \ref{tab:table1}. All items were real Facebook posts obtained via the CrowdTangle API. 

\begin{table} 
    \caption{Dimensions of content inventory}
    \centering
    \begin{adjustbox}{width=0.8\textwidth}
    \begin{tabular}{|c|c|c|}
      \hline
    \rowcolor[HTML]{F3F3F3} 
    \textbf{Dimension}       & \textbf{Low-quality}      & \textbf{High-quality} \\ \hline
    Trustworthiness & Untrustworthy sources     & Trustworthy sources   \\ \hline
    Veracity                 & Misinformation/Conspiracy & Fact-checking         \\ \hline
    Hyper-partisan & \begin{tabular}[c]{@{}c@{}}Hyper-partisan \end{tabular} & Non-political, educational \\ \hline
    
    \end{tabular}
    \end{adjustbox}
    \label{tab:table1}
    \Description{Dimensions of content inventory}
\end{table}

In Table \ref{tab:table1}, \emph{Trustworthiness} dimension includes posts from trustworthy news outlets with a NewsGuard rating of 60 or above, as well as posts from untrustworthy ones scoring below 60 on the NewsGuard ratings. (Examples of trustworthy sources: NYT, CNN, NBC News, Fox News; Untrustworthy: Breitbart, Daily Kos, Dan Bongino.) \emph{Veracity} dimension contains misinformation/conspiracy and fact-checking sources, (Examples of misinformation/conspiracy sources: Truth Theory; Fact-checking: Politifact.), while \emph{Hyper-partisan} dimension includes content from both Pro-Democrat and Pro-Republican sources, contrasted with non-political/educational content. (Examples of pro-Democrat hyperpartisan sources: The Other 98\%, Palmer Report; Pro-Republican hyperpartisan: The Daily Caller, Republican Patriots; Non-political/educational: Nature, Science.)

For mock feeds in the survey, we selected 8 social media posts, organized into contrasting pairs within each of key dimensions (1 trustworthy vs. 1 untrustworthy; 1 fact-checking vs. 1 conspiracy theory; 1 pro-Democrat, 1 pro-Republican vs. 2 non-political scientific posts). 

For interviews, the full inventory in Fig.~\ref{fig:Fig1} consisted of 15 items, identical across participants but distinct from the survey items while covering the same dimensions. The choice of 15 posts is based on a study finding that users viewed an average of 30 tweets in one session of using Twitter/X \cite{BouchaudEtAl2023}. We limited to 15 items to account for longer texts and larger images in Facebook posts, as well as to reduce cognitive load for participants. 

Additionally, network tags are added to four randomly selected posts within the full inventory. These tags represent the persona’s social network, including both strong ties (“\emph{Your friend liked}” and “\emph{Your friend reposted}”) and weak ties (“\emph{Student Union liked}” and “\emph{Student Club reposted}”). By incorporating such tags, we aimed to create a more ecologically valid representation of a social media news feed as well as explore how curation decisions shift when content carries social signals.

\subsection{Analysis}
For \hyperlink{rq1}{RQ1}, we used both survey data and interview artifacts to identify gaps between stated and revealed preferences. For \hyperlink{rq2}{RQ2}, we analyzed open-ended survey responses and interview transcripts. For \hyperlink{rq3}{RQ3}, we examined the decision-making processes of participants during feed curation activities. Finally, for \hyperlink{rq4}{RQ4}, we analyzed the design ideas generated by participants during the interviews. See Table~\ref{tab:measures-rq} for an overview of data sources and corresponding analyses.

\begin{table} 
    \caption{Overview of analytical strategies mapping each question to constructs, data sources, and results}
    \begin{adjustbox}{width=\linewidth}
    \begin{tabular}{|p{1.2cm}|p{2.2cm}|p{9.5cm}|p{1.5cm}|}
    \hline
    \rowcolor[HTML]{F3F3F3}
    \textbf{RQ} & \textbf{Concepts} & \textbf{Data Source \& Measure} & \textbf{Results}  \\
    \hline
    RQ1. What gaps? 
    & Stated vs. \newline revealed \newline preferences 
    & \emph{Survey}: Coin allocation (stated preferences) vs. engagement in mock feeds (revealed preferences); \newline \emph{Interview}: RBO and $X@k$ scores comparing the participant-curated feed (based on stated preferences) vs. engagement-maximizing feed (based on revealed preferences)
    & \emph{Survey}: Fig.~\ref{fig:Fig3},\newline
      \emph{Interview}: Figs~\ref{fig:Fig5}, ~\ref{fig:Fig6}\\  
      \hline

    RQ2. Making sense?
    & Perceptions \& \newline interpretations 
    \newline of the gap 
    & \emph{Survey}: Open-ended question asking why participants engaged with certain content differently from their stated preferences; \newline
      \emph{Interview}: Curation tasks and comparison with engagement-maximized feed to reflect on perceived gaps and rationale 
    &  Section~\ref{sec:RQ2}  \\
    \hline

    RQ3. What values?
    & Value \newline dimensions \& \newline trade-offs
    & \emph{Interview}: Participant reasoning during the feed curation task, when ranking or excluding content items based on elicited values
    &  Section~\ref{sec:RQ3}  \\
    \hline

    RQ4. Design ideas?
    & Design ideas \newline \& challenges  
    & \emph{Interview}: Post-task reflections exploring feature ideas, design challenges, and suggestions for improving the feed experience 
    &  Section~\ref{sec:RQ4}  \\
    \hline
    \end{tabular}
    \end{adjustbox}
    \label{tab:measures-rq}
\end{table}

\subsubsection{Survey}
We first compared the distribution of engagement behaviors (revealed preferences) and coin allocations (stated preferences) across source types to see whether and where gaps manifest (\hyperlink{rq1}{RQ1}). Second, we qualitatively analyzed open-ended responses to explore the perceptions of participants of the gaps and their mental models of social media algorithms (\hyperlink{rq2}{RQ2}).

\subsubsection{Interview with feed curation activities}
The interviews allowed us to address all four questions: we analyzed participant-curated feeds alongside the engagement-maximizing feed to assess how stated preferences diverged from typical algorithmic outputs (\hyperlink{rq1}{RQ1}); explored how participants made sense of identified gaps (\hyperlink{rq2}{RQ2}); examined what values they considered when envisioning an ideal news feed and how they navigated trade-offs among competing priorities (\hyperlink{rq3}{RQ3}); and identified design principles and features for more value-aligned recommendation systems (\hyperlink{rq4}{RQ4}).

Specifically, for RQ1, we compared participant-curated feeds against the engagement-maximizing feed using two metrics: Ranked-Biased Overlap (RBO) and Precision@k. RBO measures similarity between two feed rankings \cite{WebberEtAl2010}, with higher RBO scores indicating greater similarity. A low RBO score was expected between participant-curated and engagement-maximizing feeds, indicating a gap between stated preference-based and revealed preference-based ranking approaches. To assess statistical significance, a null distribution of RBO scores was generated by comparing randomly simulated feeds against the engagement-maximizing feed.

In the same spirit of the Precision@$k$ metric \cite{ZangerleEtAl2023}, we adopted a standard practice in the recommender system field to calculate $X@k$ metrics as the proportion of items with a specific characteristic $X$ within the top $k$ ranked items. While RBO measures overall similarity, the following $X@k$ metrics show how feeds prioritize specific content characteristics. 
\begin{itemize}
    \item Ideological Slant@$k$ measures political leanings of a given feed. We assigned ideology scores to posts ($-1$ for pro-democrat, $0$ for neutral, $+1$ for pro-republican) based on classifications of source bias provided by Media Bias/Fact Check (MBFC; https://mediabiasfactcheck.com/). We then calculated the average ideology score of top $k$ ranked posts, where a positive score indicates a conservative-leaning, while a negative score suggests a liberal-leaning feed. 
    \item Cross-Cutting Exposure@$k$ ranges from $0$ (one ideology) to $1$ (a perfectly balanced mix), with higher values indicating more cross-cutting exposure and lower values implying ideological homogeneity of top $k$ posts. 
    \item Trustworthiness@$k$ calculates the average NewsGuard scores within top $k$ posts, ranging from $0$ to $100$, while Credibility@$k$ calculates the average credibility score of sources in top $k$ posts using MBFC’s ratings, where each source is assigned a weight ($1$ for high, $0.5$ for medium, $0$ for low credibility). For both metrics, higher values indicate higher quality.
\end{itemize}

For RQ2--RQ4, an initial codebook was developed based on the research questions and interview protocol \cite{saldana2021coding}, resulting in 8 themes and 17 sub-codes. Two researchers independently coded the same randomly selected transcript using NVivo. After the initial round, the codebook was iteratively refined by merging overlapping codes, clarifying definitions, and adding newly emergent ones, which resulted in a total of 8 themes and 20 sub-codes. Next, the team independently coded another randomly selected transcript using the revised codebook, iterating on further refinements. The final codebook (See Appendix~\ref{appendix:codebook}), with 8 themes and 30 sub-codes, was then applied to the full dataset.

\section{\MakeUppercase{Results}}

\subsection{What Gaps Exist Between Revealed and Stated Preferences?}

\begin{figure} 
    \includegraphics[width=\textwidth]{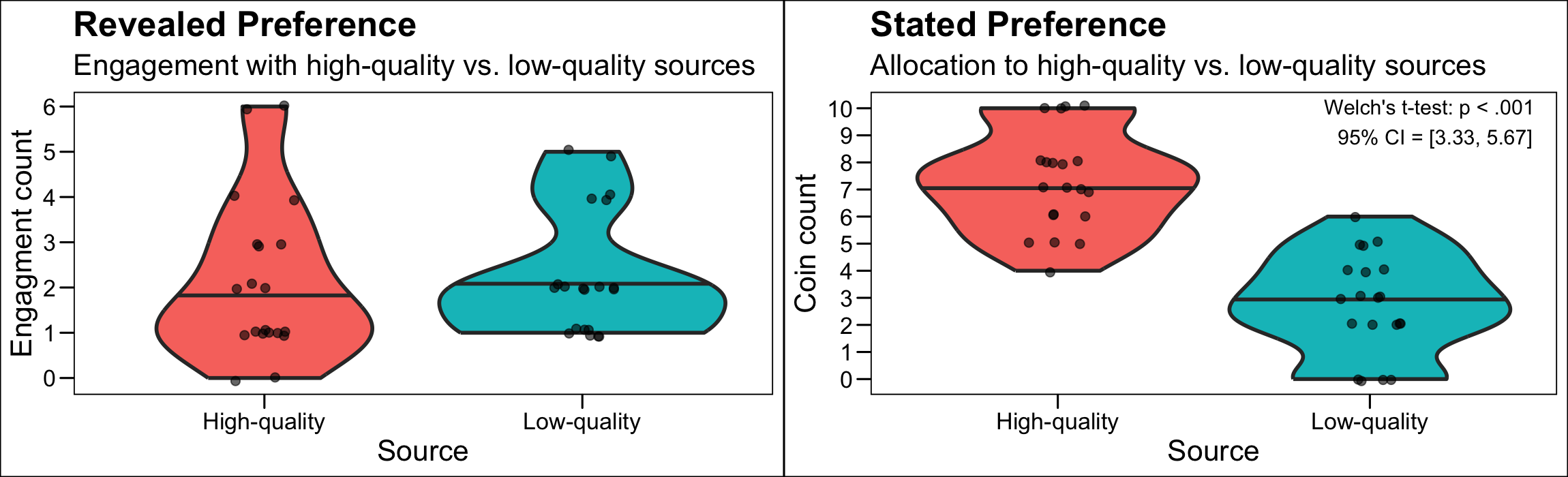}
    \centering
    \Description{Figure 4. Revealed vs. stated preferences for high- and low-quality sources.}
    \caption{Violin plots of revealed vs. stated preferences for high- and low-quality sources, measured in the survey: The left panel shows engagement count as a measure of revealed preferences, while the right panel shows the number of coins allocated as a measure of stated preferences. Horizontal lines are the median.}
    \label{fig:Fig3}
    \vspace{-0.5em}
\end{figure}

While participants explicitly preferred high-quality sources, their engagement patterns showed no such preference. Fig.~\ref{fig:Fig3} illustrates this gap, with Welch's t-test results reported only for comparisons with significant differences. As shown in the left panel, participants did not necessarily engage more with high-quality sources than with low-quality ones, whereas the right panel shows that they allocated significantly more coins to high-quality sources.

\begin{figure} 
    \includegraphics[width=0.9\textwidth]{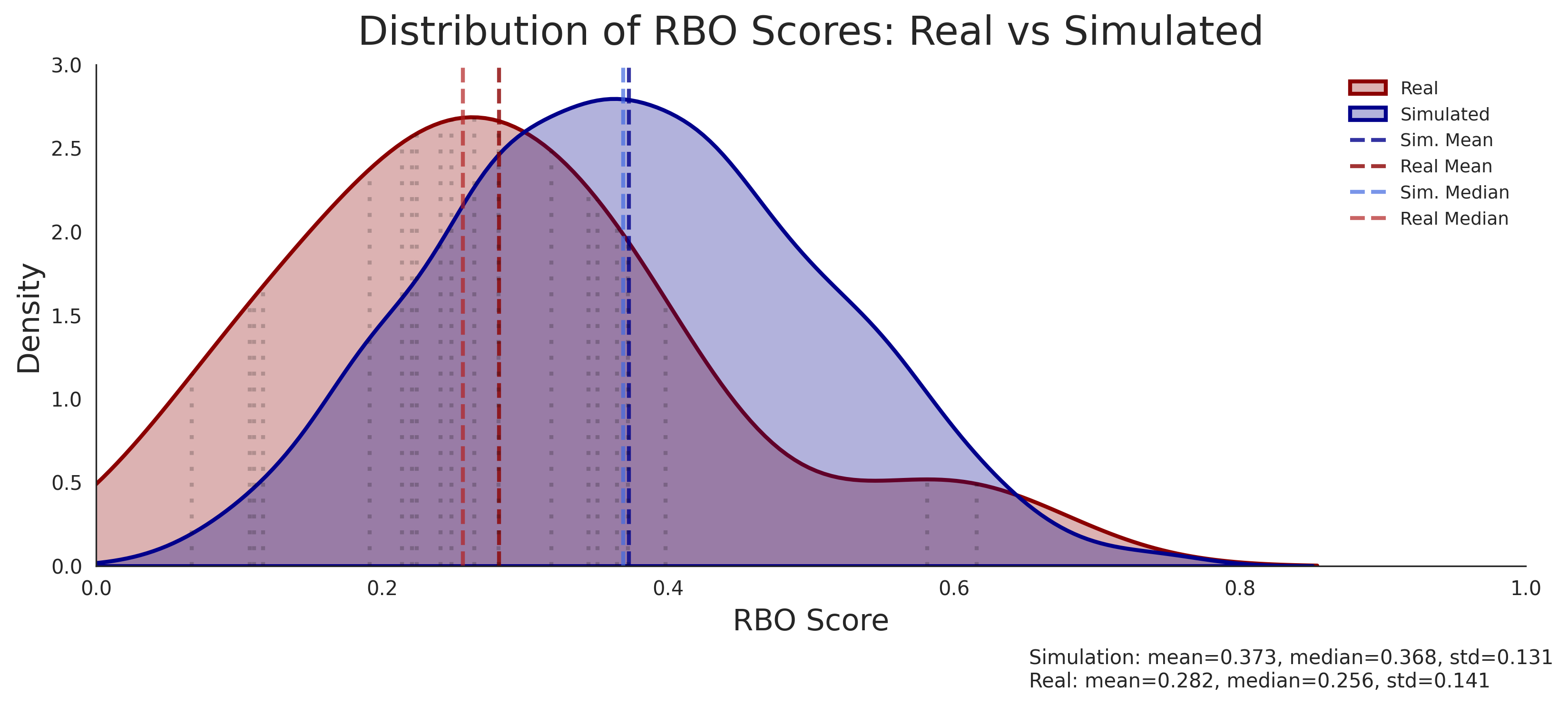}
    \centering
    \Description{Figure 5. Distribution of RBO scores for real user-curated feeds and randomly simulated feeds, compared to the engagement-maximizing feed.}
    \caption{Density plots showing the distribution of RBO scores for real user-curated feeds (in red) and randomly simulated feeds (in blue) compared to the engagement-maximizing feed. Higher RBO scores indicate a greater alignment with the engagement-maximizing feed.}
    \label{fig:Fig5}
\end{figure}

Fig. \ref{fig:Fig5} shows the distributions of RBO scores (1) between the user-curated feeds and the engagement-maximizing feed, and (2) that between randomly generated feeds and the engagement-maximizing feed. The random feeds were created by randomly sampling items within the same content inventory, with each generated feed matched to the average length of the user-curated feeds.  
The average RBO score between curated feeds and the engagement-maximizing feed was significantly lower than that between the randomly generated feeds and the engagement-maximizing feed ($t = -2.819$, $p = 0.01$, Cohen's $d = -0.669$). Thus, stated preference-based feeds diverged from the engagement-maximizing feed even more than a purely random feed did. 

\begin{figure} 
    \includegraphics[width=0.9\textwidth]{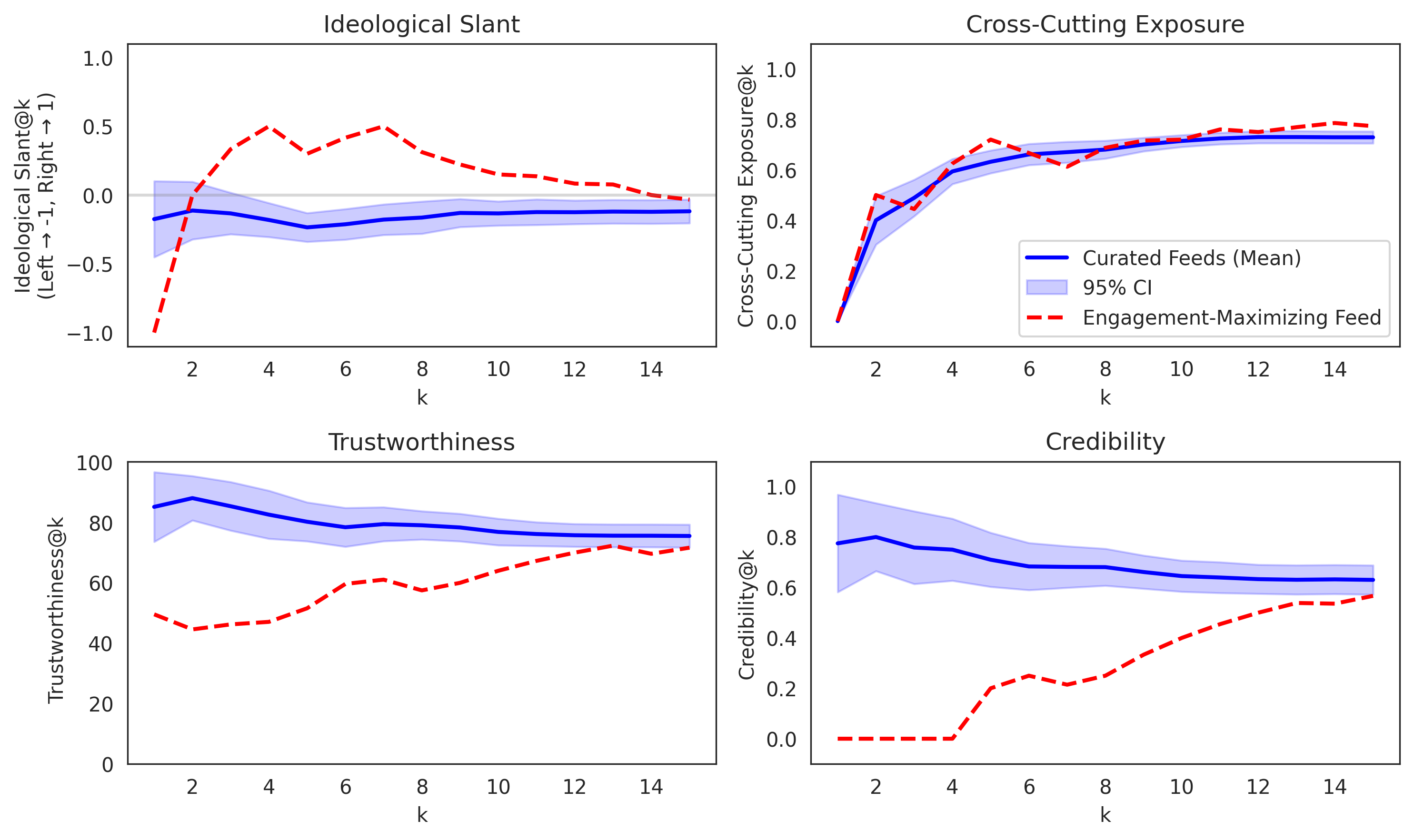}
    \centering
    \Description{Figure 6. Comparison of attribute scores between the engagement-maximizing feed and user-curated feeds.}
    \caption{Comparison of feed characteristics: The x-axis represents the ranking position k, while the y-axis shows attribute scores at each k. Blue lines represent average attribute scores for curated feeds, with a 95\% confidence interval, and red dashed lines are the attribute scores for the engagement-maximizing feed.}
    \label{fig:Fig6}
\end{figure}

Fig. \ref{fig:Fig6} shows how user-curated feeds and the engagement-maximizing feed prioritized different attributes across rankings. In the top-left panel (\emph{Ideological Slant}), the engagement-maximizing feed generally leans to the right. In contrast, the curated feeds lean to the left, which makes sense given the political leanings of recruited participants (see Table~\ref{table:table5}).

In the top-right panel (\emph{Cross-Cutting Exposure}), the engagement-maximizing feed presents a mix of both right- and left-leaning sources. This is because the full inventory used in this study was deliberately designed to include various political perspectives. Since the algorithm reorders this mixed inventory based on engagement metrics, the feed naturally features content from across the ideological spectrum, inherently providing a degree of cross-cutting exposure. 

Interestingly, participant-curated feeds show a similar level of cross-cutting exposure, suggesting that optimizing for the stated preferences of users does not necessarily lead to echo chambers, as some might fear \cite{MilliEtAl2023a}. These concerns often arise from the idea that, if users are only exposed to content that they want to see, they may end up seeing only viewpoints that agree with their own. However, participants stated a desire for exposure to politically diverse perspectives. While this may relate to the fact that 30\% of interviewees were independent, these findings suggest that stated preference-based ranking could naturally encourage exposure to a broader range of viewpoints. 

Finally, the bottom panels in Fig. \ref{fig:Fig6} show that participant-curated feeds predominantly feature trustworthy and credible sources. Although two feeds converge as the ranking extends, which is expected given that both feeds are based on the same, limited pool of content, the sharp contrast at the top $k$ indicates the engagement-maximizing feed prioritizes engagement over trustworthiness or credibility. This suggests that stated preference-based ranking methods may offer a better alternative, preserving feed quality while aligning more closely with the stated preferences of users.

\subsection{How Do Users Understand and Interpret These Gaps?}\label{sec:RQ2}
Having reaffirmed the well-documented gaps between stated and revealed preferences in our sample, we next examined how participants made sense of these gaps. We analyzed the initial mental models of participants about algorithms and their experiences of the gap (\emph{4.2.1}), how they interpreted the gap when comparing the curated and engagement-maximizing feeds (\emph{4.2.2}), and how their justifications for why such gaps exist (\emph{4.2.3}).

\subsubsection{Initial mental models of algorithms and experience of the gap}
Mental models or folk theories of algorithms of users shape how they interact with these systems \cite{StormsEtAl2022, EslamiEtAl2016, KarizatEtAl2021}. Thus, we examined their understanding of social media algorithms and their experiences regarding the gap between revealed and stated preferences. Analysis of survey responses revealed that participants believed algorithms primarily rely on engagement patterns. P7 said that the feed showed the content they “\emph{interacted with most},” and P3 thought “\emph{search history}” played a role. 

Nonetheless, participants questioned whether such algorithms truly captured their genuine preferences. P6 mentioned that liking a conspiracy theory post once resulted in an unwanted flood of similar content, emphasizing that engaging with a post does not always imply a desire for more of the same kind of content. Participants also noticed that factors unrelated to their preferences, such as popularity (P5) or advertising (P10 and P12), may influence algorithms.

Interestingly, satisfaction with news feeds seemed closely tied to how much control participants felt they had. Those who reported being ‘most satisfied’ with their current news feeds also recognized their agency in content curation, noting: “\emph{I and I alone curate my own feed and content}” (P10). Dissatisfied users, on the other hand, felt that algorithms operated with an agenda irrelevant to their preferences. For instance, P12 noted, “\emph{The platform pushes popular stuff that isn’t necessarily what I want}.” This is reminiscent of the gap in (lack of) trust in competence (i.e., the ability of an algorithm to identify relevant information) versus (lack of) trust in motivation (i.e., the objectives behind the design of an algorithm) that is observed in risk communication \cite{Twyman_Harvey_Harries_2008}. 

\subsubsection{Making sense of the gap: Comparing user-curated and engagement-maximizing feeds}

Table \ref{tab:table2} summarizes differences participants noted between curated and engagement-maximizing feeds. P10 noted the engagement-maximizing feed was “\emph{more about shock value and entertainment},” while P1 thought “\emph{it’s trying to generate controversy to drive up engagement}.” Curated feeds, on the other hand, were more aligned with what participants valued, being “\emph{factual and reputable rather than engagement-driven}” (P12). These distinct priorities led to differences in key aspects of each feed, such as feed composition, the format and style of promoted posts, and the overall user experience.

\begin{table} 
    \caption{Comparison of the engagement-maximizing feed and participant-curated feeds}
    \begin{adjustbox}{width=\textwidth}
    \begin{tabular}{|c|c|c|}
    \hline
    \rowcolor[HTML]{F3F3F3} 
    \multicolumn{1}{|l|}{\cellcolor[HTML]{F3F3F3}\textbf{}} & \textbf{Engagement-Maximizing Feed} & \textbf{Curated Feeds}    \\ \hline
    Core priorities                                & Max(Engagement)                     & Max(Values)               \\ \hline
    Feed Composition                                        & Biased; exposure to one-sided views & Exposure to diverse views \\ \hline
    Content format                                          & Shorter, meme/screenshot images     & Longer, text-based        \\ \hline
    User experience &
      \begin{tabular}[c]{@{}c@{}}Less satisfied: Divisive engagement; \\ oversimplified\end{tabular} &
      \begin{tabular}[c]{@{}c@{}}More satisfied: Healthy engagement; \\ more personalized\end{tabular} \\ \hline
    \end{tabular}
    \end{adjustbox}
    \label{tab:table2}
\end{table}

Participants saw the composition of the engagement-maximizing feed as biased. P1 said “\emph{You don’t really see another side, or even an alternative perspective},” while P6 said it contained “\emph{more polarizing content\ldots before getting into anything more balanced}.” By contrast, user-curated feeds encouraged exposure to diverse viewpoints. As P6 explained, “\emph{my feed was balanced to show both sides and encourage the [persona] to form their own opinions}.” 

Moreover, the engagement-maximizing feed prioritized short posts as P10 noted, “\emph{content at the top is brief and visual... like memes or shocking images}.” In contrast, curated feeds had longer, educational posts that were less engaging, as P12 observed, “\emph{informative posts are at the bottom because they get less engagement}.” 

Overall, participants were more satisfied with curated feeds than the engagement-maximizing feed, though there were some exceptions (see Section~\ref{sec:exceptions}). Some found the engagement-maximizing feed overwhelming and stressful, describing it as a “\emph{war zone}” (P4), “\emph{tearing everybody apart}” (P3), or encouraging “\emph{negative behaviors being targeted toward other groups}” (P1). Conversely, participants perceived the curated feed as a user-centered approach that encouraged thoughtful engagement. 

Some participants even pointed out that the engagement-maximizing feed oversimplifies users, reducing them to narrow profiles derived from sparse interactions. P6 explained that it “\emph{strips users of their individuality and limits exposure to diverse perspectives},” while P8 remarked that users “\emph{are no longer themselves; they're literally their news feed},” making it difficult to separate personal beliefs from algorithmically reinforced ones.  

\subsubsection{Justifications for the gap}
Participants reasoned that these observed differences stem from multiple factors: (\emph{i}) controversial content attracts more engagement, (\emph{ii}) generating revenue for platforms, and (\emph{iii}) highly active users engage more, causing algorithms to disproportionately amplify their voices. 

Participants reasoned that the engagement-maximizing feed algorithm promoted provocative content, even when such content contradicts what users truly prefer to see. P1 noted, “\emph{It’s more like generating rage and controversy to drive up likes and engagement}.” 

They also attributed this emphasis on engagement-maximizing content to the financial incentives of the platforms. As P11 speculated, “\emph{I wonder if there’s money behind that},” implying that platforms may prioritize content that keeps users engaged to generate revenue, potentially over content aligned with users’ values.  

Participants further noted that engagement-driven algorithms can push controversial or divisive content to the top---even if it is not what they truly prefer---because highly opinionated users interact more frequently and intensely with content, leading algorithms to boost their voices disproportionately. P3 further suggested that content with both sides of the political spectrum can generate even more interaction: “\emph{It’s probably at the top because it has both sides of the party\ldots because they’re fighting with each other}.” 

\subsection{What Value Dimensions Do Users Consider, and How Do They Navigate Tensions Between Them?}\label{sec:RQ3}
In this section, we examine the specific value dimensions that emerged from the interviews (\emph{4.3.1}), explore how participants navigated tensions between competing values (\emph{4.3.2}), and investigate how contextual factors shaped their decisions (\emph{4.3.3}). These insights not only help explain why participants favored their curated feeds over the engagement-maximizing feed but also suggest principles for designing value-aligned algorithms.

\subsubsection{Value dimensions}

\begin{table}
    \centering
    \caption{Key value dimensions for ideal news feeds, with number of participants who mentioned each value.}
    \begin{adjustbox}{width=\textwidth}
    \begin{tabular}{|c|m{0.65\textwidth}|c|}
        \hline
        \rowcolor[HTML]{F3F3F3} 
        \textbf{Value dimensions} & 
        \multicolumn{1}{c|}{\cellcolor[HTML]{F3F3F3}\textbf{Description}} & 
        \textbf{N} \\ \hline
                
        \textbf{\begin{tabular}[c]{@{}c@{}}Balanced/Diversity \\(B/D)\end{tabular}} & 
        Ensuring diverse perspectives rather than one-sided bias. 
        \emph{Example: Including news from both left- and right-leaning sources.} & 16 \\ \hline

        \textbf{\begin{tabular}[c]{@{}c@{}}Trustworthy/Accuracy \\(T/A)\end{tabular}} & 
        Prioritizing credible and fact-based content over untrustworthy sources. 
        \emph{Example: Ranking fact-checked or science-based articles higher in the feed.} & 11 \\ \hline

        \textbf{\begin{tabular}[c]{@{}c@{}}Sensitivity/Ethical \\(S/E)\end{tabular}} & 
        Avoiding inflammatory or overly negative content for ethical reasons.
        \emph{Example: Filtering out posts that incite fear or hate.} & 9 \\ \hline

        \textbf{\begin{tabular}[c]{@{}c@{}}Informative/Educational \\(I/E)\end{tabular}} & 
        Content with meaningful information or learning value. 
        \emph{Example: Promoting educational articles on science, history, or policy.} & 9 \\ \hline

        \textbf{\begin{tabular}[c]{@{}c@{}}Relevance \\(REL)\end{tabular}} & 
        Topics that are timely or widely discussed. 
        \emph{Example: Including posts that are widely shared or deal with major current events.} & 7 \\ \hline

        \textbf{\begin{tabular}[c]{@{}c@{}}Entertaining \\(ENT)\end{tabular}} & 
        Including content that is engaging and enjoyable. 
        \emph{Example: Featuring viral memes or content that is fun to see.} & 5 \\ \hline
    \end{tabular}
    \end{adjustbox}
    \label{tab:table3}
\end{table}

Six value dimensions emerged as guiding principles for how participants curated their ideal social media feeds (see Table \ref{tab:table3}).
Most discussions centered on Balance/Diversity (B/D) and Trustworthy/Accuracy (T/A), reflecting a desire for both multiple perspectives and reliable sources. Participants also considered Sensitivity/Ethical (S/E) and Informative/Educational (I/E) values, while Relevance (REL) and Entertaining (ENT) received less emphasis. While current algorithms largely prioritize relevance and entertainment, participants emphasized social values such as diversity, accuracy, and ethical considerations.

When dumping or ranking posts lower in their feeds, perceived bias (B/D) of the content was a common reason, as P1 dumped a post “\emph{because it’s kind of biased towards a particular political party}.” Another reason for dumping was perceived lack of credibility (T/A). P6 shared that “\emph{I feel like this is a little conspiratorial},” as well as P11: “\emph{This post mirrors a lot of fake posts that I’ve seen}.” Content that lacked substance or of little informational value (I/E) was also downranked, including posts that “\emph{don’t really add anything}” (P12). Participants excluded content that was irrelevant (REL) (“\emph{I don’t know if they would care about that}", P10) or overly negative (S/E) (“\emph{I don’t want to dive too much into darker topics\ldots I don’t think they need to deal with something like this}", P6). 

When ranking posts higher, creating a feed with multiple or balanced perspectives (B/D) was especially valued. For instance, P2 put “\emph{Fox News and CNN next to each other... just so you’re not being constantly bombarded with one kind of content}.” Similarly, P13 stressed the importance of the chosen person being exposed to the opposing viewpoints, “\emph{not to be stuck in a preconception that they’re the good guys and can do no bad}.” Factual accuracy or credibility (T/A) were also crucial, as participants noting factual posts were “\emph{very useful}” (P12). Likewise, educational (I/E) content was valued as well. Furthermore, content that resonated with popular (REL) opinion was ranked higher, as P10 emphasized, “\emph{it’s something that the majority of Americans agree with},” while P11 remarked, “\emph{it’s good for the [persona] to consider what the general opinion is}.”

\subsubsection{Tensions between values} 
Participants wanted to make sure their curated feeds were balanced and trustworthy, but they also needed to keep them interesting and relevant. As a result, they negotiated trade-offs between their stated values and more engagement-oriented considerations.

We found conflicts between prioritizing accuracy or credibility and entertaining content. In many cases, participants prioritized accuracy or credibility, even when this meant potentially reducing engagement. For instance, P1 noted: “\emph{there can be entertainment mixed into it. But I feel like the facts come first and the trustworthiness}.” 

Some participants found middle ground to include engaging content in their feeds without compromising their values. P8 showed this balance by saying, “\emph{the [persona] would enjoy a bit of entertainment\ldots also they would want to know what is going on}.” While acknowledging the importance of entertaining and relevant content, P8 only allowed such content up to a point, noting “\emph{throwing insults back and forth; that’s where I draw the line}.” 

The tension between values attached to engagement maximization and other values was also evident when participants tried to prioritize diversity. For example, despite personally disliking content “\emph{trying to say you’re wrong},” P3 included such content only when it served their stated value of enabling users to explore different perspectives, saying “\emph{if they are independent and want to see both sides of things}.”

Meanwhile, P11 faced a dilemma when deciding whether to include content that, although factually accurate, could promote division between political groups: “\emph{I was debating whether or not to throw this out, because I think it is promoting negative vitriol between the parties. But at the same time, I think it does adhere to factual accuracy}.” This example reflects how participants had to weigh their core value such as accuracy against their desire to create a less polarizing environment, illustrating the trade-offs involved in maintaining multiple quality standards simultaneously.

\subsubsection{Decision-making context}
Various contextual factors played a role in the curating process: (\emph{i}) their assigned persona; (\emph{ii}) any social cues attached to the content; (\emph{iii}) their partisan identification; and (\emph{iv}) their prior knowledge of the sources of the content. 

First, participants considered the traits and preferences of personas they could easily relate to. As P12 stated, “\emph{I feel similar. I feel like I kind of understand this person}.” However, participants also considered distinct values of their personas. P11, for example, tailored content for the chosen persona: “\emph{But just because I am a liberal\ldots since I’m curating this thing for someone who’s like a [persona], I’m not going to exclude things just because it’s by a conservative news source}.” 

Second, participants were more likely to include or prioritize content liked or shared by someone within the persona's network. P10 said, “\emph{If their friend reposted this, then I guess they might also try to}.” As P12 noted, participants “\emph{put things that people that your friend or similar groups liked above random things}.” This suggests that content endorsed by social ties held particular weight in their curation decisions. 

Third, although partisan predispositions of participants affected their curating decisions, they also tried to mitigate their partisan biases. P11 admitted, “\emph{My own biases as a liberal person might be factoring into me saying that}.” P7 shared a similar struggle: “\emph{I don’t agree with most of the things [featured politician in the post] says. I try to keep my personal political views out of it, though}.” 

Lastly, participants actively utilized their prior knowledge of sources when curating the feed. For example, P12 expressed confidence in upranking the fact-checking source, which “\emph{is known to be very reputable}.” However, this reliance on background knowledge also meant participants needed alternative signals if they were unfamiliar with a source. As P18 explained, “\emph{I don’t know how reliable [the source] is, but at least it provides a link for further reading}.”

Overall, feed curation is a form of socially embedded judgment. Participants interpreted what the persona should see not just in terms of interest or relevance, but in relation to perceived norms, social cues, and broader questions about what is appropriate in a shared information environment.
 
\subsection{Exceptions: Participants Who Emulated the Engagement-Driven Algorithms}\label{sec:exceptions}
A subset of participants (N=4) mirrored current social media recommendation algorithms. First, P17 and P20 deliberately approached the feed curation task from the perspective of a social media company. P20, a self-identified independent who viewed social media as a source of fun rather than news, framed the task as an exercise in keeping people on the platform. P20 explained, \emph{“So if I’m a company, my goal is to get them to stay on the website as long as possible.”} P17, a self-identified Republican, emphasized the potential for controversy to captivate users: \emph{“if we want the user to really actually dive into the feed, then we want something that’s unusual for them to see.”}  

A different kind of exception arose with P14 and P16, both self-identified Republicans with limited familiarity with U.S. political contexts and the news sources provided. Although they wanted to create “\emph{educational}” and “\emph{trustworthy}” feeds, their lack of knowledge about reliable versus unreliable outlets made it difficult to achieve this goal. Lacking sufficient background information, these well-intentioned participants inadvertently curated content much like an engagement-driven algorithm would, favoring controversial or attention-grabbing posts.  

These exceptions hint at the complex interplay of information ecosystems, individual values, and political predispositions. Among Republicans, this dynamic sometimes manifested intentionally (e.g., P17) and other times inadvertently (e.g., P14, P16), reflecting an alignment with algorithmic logic embedded in social media platforms. Similarly, independents like P20 prioritized entertainment over other values, further showing how political orientation---or the absence of it---shapes feed curation approaches. We leave to future research a deeper examination of how political predispositions shape feed curation strategies.

\subsection{What Design Features Do Users Suggest for Value-Aligned Recommendation Systems?}\label{sec:RQ4}
This section presents five proposals for value-aligned algorithms along with challenges and opportunities in implementing these designs. 

\subsubsection{Proposed design}
\begin{enumerate}
    \item Participants emphasized the need for recommendation systems to better incorporate stated preferences of users through both initial setup and ongoing controls. P10 suggested collecting user preferences as an initial step. To ensure feeds continue reflecting evolving preferences during use, P12 envisioned a sliding scale on the feed, with which users “\emph{can set a temperature of how varied they want their feed to be}.” 
    
    \item Participants called for more accessible and effective controls over their feeds, highlighting how existing tools are often insufficient or difficult to use. P6 proposed giving users more controls, but also noted that existing tools are often hidden in settings, making them hard to use effectively: “\emph{If we made those options more accessible, they’d have more control over what they see}.” 

    \item Participants also proposed the implementation of customizable algorithmic modes, including switching between different recommendation systems or adjusting specific sorting parameters. P6 suggested “\emph{giving users the ability to switch between different algorithms}.” P12 and P18 made similar suggestions: “\emph{Sorting algorithms would be good. Right now you can only sort by recency or post count},” and features that allow users to “\emph{sort by reliability}.”  
    
    \item Regarding algorithmic transparency, participants wanted a clearer sense of how content is shown to them. 
    They also proposed features such as built-in fact-checking tools that would allow them to instantly see if content is accurate. P3 emphasized this point, stating, “\emph{I wish I had proof that it was actually fact-checked}.” 
    
    \item Finally, participants observed that users also need to develop their own media literacy skills, as P6 stated: “\emph{I don’t think a media feed should necessarily tell them what’s right or wrong; it should be up to them to discern that}.” Interestingly, the interview itself served as an algorithmic literacy exercise. For P7, for instance, the interview led to “\emph{realize that what’s on my ‘For You’ page isn’t always something I care about or agree with}.” These findings are consistent with the `bi-directional alignment,' in which AI systems are designed to align with human values while humans are also empowered to better articulate their values \cite{ShenEtAl2024}. 
\end{enumerate}

In summary, there is a need for better ways to reflect the stated preferences of users and allow them to actively shape their feeds with greater control and customizable options. Transparency and algorithmic literacy serve as complementary measures, empowering users to make informed choices and critically engage with content. Together, these design recommendations contribute to the goal of building a more user-centered, value-aligned recommendation system. 

\subsubsection{What are the perceived challenges and opportunities in implementing these proposed ideas?}
Platform incentives are a commonly perceived challenge. Participants noted that platforms prioritize maximizing engagement for revenue, which may be in conflict with prioritizing user values, as P11 observed: “\emph{I don’t know if it’s necessarily the most financially advantageous thing to do}.” 

Another challenge relates to the fact that user preferences may not be internally consistent nor be completely formed, leading to idiosyncratic news consumption habits. P10 explained, “\emph{people might not care},” while P11 noted that many users “\emph{don’t have a strong preference}.” Even for those who do care and have clear preferences, as P19 put it, “\emph{some days they might be feeling liberal, other days something else.}” This implies that a system must either update constantly or rely on frequent user input to keep pace with changing preferences. 

Overall, participants anticipated that a key challenge might be motivating both platforms and users to look beyond immediate engagement and prioritize long-term quality and well-being. Even the best-designed interventions will fall short unless platforms see an incentive to implement them and users find value in devoting their time and effort. Without a clear path for sustaining revenue and user interest in higher-quality experiences, any improvements remain merely aspirational.

Despite these challenges, there were also opportunities. One idea is to adopt an alternative business model that moves away from engagement-driven incentives. P2 suggested a paid tier system: “\emph{Having like a paid tier model for the platform...you can have better features, or like you can have an upgraded experience if you pay for this}.” Such a model could sustain the platform financially while providing enhanced user control. Similarly, P18 suggested incentivizing advertisers to place ads alongside reliable content rather than popular but untrustworthy posts. This echos with a finding where many corporate decision-makers were unaware their ads ran on misinformation websites, yet strongly preferred not to fund such websites once they realized it \cite{AhmadEtAl2024}. 

Participants also recognized an opportunity for aligning long-term profit motives and user needs. By focusing on the values users prioritize, a platform could increase user engagement, thus supporting its own incentives, through fostering trust in the platform. P5 mentioned, “\emph{I feel like I would trust that platform more than one that isn’t using those values}.” 
Some further suggested that value-aligned recommendation systems could enhance mutual understanding and bridge social divides. P6 imagined that “\emph{it would help people become more open-minded\ldots If the algorithm exposes people to a wider range of opinions, it might reduce division and encourage more dialogue}.”

\section{\MakeUppercase{Discussion}}
Prior work has documented the existence of a gap between stated and revealed preferences in social media users. By focusing on the social media experience of U.S.-based young adults through a mix of surveys and feed curation activities, here we extend this literature by showing that a stated preference-based ranking algorithm can serve as a promising complement to standard engagement-based ones, with the potential to improve feed quality and user satisfaction. We also show how participants understand this gap through the feed curation task. Participant reflections suggest that engagement metrics persist in practice not because they reflect the underlying values of users, but because they align with broader platform incentives. From this perspective, the gap between stated and revealed preferences reflects the fact that social media are multi-sided markets \cite{Abdollahpouri2022, rochet2003platform}, featuring the interplay between economic incentives underlying engagement-optimized algorithmic systems and the constrained capacity of users to meaningfully intervene in these systems.

While previous research has called for a greater incorporation of stated preferences into algorithmic design~\cite{StrayEtAl2023, MorewedgeEtAl2023}, how to elicit and operationalize them in dynamic, content-rich environments has remained an open question. This study addresses that challenge by structuring a task where users design news feeds for imagined personas based on their stated values. We identified recurring value dimensions participants think should be prioritized in feed design, and surfaced how they reasoned through trade-offs and translated these values into concrete ranking choices. Furthermore, in an ideation task, participants proposed a range of design ideas to better align algorithmic curation with their stated values, emphasizing the need for interfaces that make values visible, negotiable, and actionable. 

We extend CSCW and HCI research on value-sensitive and user-centered design~\cite{FengEtAl2024, LeeBaykal2017, ZhuEtAl2018, EslamiEtAl2016, ParkEtAl2022, KarizatEtAl2021, Swart2021, NguyenEtAl2024, StormsEtAl2022} by offering an account of how value alignment can be practically supported in algorithmic environments through participatory mechanisms. In doing so, our study explores design opportunities as well as open challenges for advancing more adaptive and accountable systems in dynamic, algorithmically mediated environments. 

\subsection{Inferred Design Features}
Before proposing concrete design pathways (see below), we distill a set of desirable features for a value-aligned news feed, grounded in our empirical findings and prior literature.

First, systems must support users in articulating what they value. Our results show that participants clearly distinguished between content that tends to draw more engagement and content they believed should be prioritized in an ideal feed, echoing prior work that critiques treating engagement as a proxy for reflective or long-term preferences \cite{MorewedgeEtAl2023, StrayEtAl2023}.

Second, value alignment actions should operate at the entire feed level rather than relying exclusively on item-level feedback, as emphasized in prior CSCW research \cite{StrayEtAl2023}. Participants reasoned about their feeds holistically (e.g., weighing the diversity and balance of perspectives at the feed level) instead of evaluating individual posts in isolation.

Third, systems must make visible and negotiable these trade-offs among competing values. Participants consistently navigated tensions between values such as trustworthiness and relevance. Rather than seeking to optimize a single value, they treated feed curation as a process of balancing competing priorities.

Last but not least, the system should support user reflection, rather than merely offering control mechanisms. Consistent with prior CSCW research~\cite{ShenEtAl2024, StrayEtAl2023, Gabriel2020, LeeEtAl2019}, we see that when curating feeds for personas, participants drew on socially grounded values to make situated decisions that reflected broader concerns about what others should see and what content is worth amplifying or avoiding. Building on this line of work, we provide a few suggestions for potential design pathways below.

\subsection{Design Pathways for Stated Preference–Based Feed}
How can we practically incorporate user values into feed algorithms? 
Participants suggested a few UI components, including slider-based interfaces that allow users to adjust feed-level value dimensions. Slider-based controllers have previously been explored as a way of increasing user agency over algorithmic feeds. Harambam et al.~\cite{HarambamEtAl} introduce sliders that allow users to adjust the relative importance of predefined topic categories in news feeds, while systems such as Gobo~\cite{Gobo} provide sliders to adjust predefined dimensions (e.g., seriousness, rudeness, or politics). Similarly, we posit that providing slider controllers corresponding to the key value dimensions identified in this study may enable users to intuitively fine-tune their feeds.  

However, a slider-based approach is not without limitations. The value dimensions themselves may evolve over time, making fixed sliders potentially outdated as user norms shift. Also, the participants in this study are not representative of all social media users, nor even of the young adult group specifically. Different populations may prioritize different values, and users may not always find their priorities fully captured by these predefined value dimensions. As these dimensions embedded in the slider-based system carry embedded judgments about what a “good” feed should look like, it could nudge users toward values preferred by designers rather than those they might have articulated themselves. This implies the need for a more adaptive and personalized approach.
 
A second approach to incorporating stated preferences into algorithmic design involves leveraging Large Language Models (LLMs) to allow users to define what they want. While participants did not explicitly suggest natural language input, their curated feeds reflected diverse interpretations of what constitutes a “good” feed. Although there were recurring value dimensions they prioritized, participants differed in how they weighed these values and what trade-offs they were willing to make. Unlike sliders, which constrain users to a preconfigured set of values, an LLM-based interface could elicit the stated preferences of users in natural language and translate them into ranking outputs. Of course, developing an LLM-based system that is stable and efficient remains an important direction for future research. 

\subsection{Challenges in Implementing Stated Preference-Based Feeds}
While stated preference-based feeds offer promising alternatives to engagement-driven feeds, they also come with challenges. Not all users have the ability to clearly articulate what they truly want. Articulating stated preferences can be cognitively demanding, and some users may struggle to define their ideal feed composition. Therefore, future research should focus on designing effective mechanisms for eliciting stated preferences, ensuring that users can express their values in a way that is both intuitive and actionable for recommendation systems.

Similarly, not all users can distinguish content that aligns with their stated preferences. Some participants in this study intended to curate value-driven feeds but found it difficult to identify content that matched their values. This suggests that, rather than placing the full burden on users to curate their feeds, the system itself should play an active role in guiding users through the process. 

These challenges do not negate the benefits of stated preference-based feeds. The very act of adjusting sliders or interacting with an LLM-powered system can help users understand how their stated preferences influence content ranking, enhancing transparency and user agency. For example, if a user requests “I want to see more factual content,” and the feed updates in real time, it offers clear feedback on how stated preferences influence recommendations. As users interact with the system, their stated preferences can be further refined through iterative feedback loops, allowing them to adjust their feed dynamically as their values evolve. 

This ties into what Eslami et al.~\cite{EslamiEtAl2016} call “seamful design,” where algorithmic processes are intentionally exposed to users rather than hidden in a black box. By showing users how adjustments such as increasing “trustworthiness” or interacting with LLM-based systems dynamically reshape their feeds, these approaches promote a more transparent, user-driven alignment between stated preferences and algorithmic outcomes. 

In addition to user-side challenges, value-aligned systems may face resistance at the platform level. While these systems may increase user trust and satisfaction, they may also reduce short-term engagement metrics that many platforms currently optimize for. Future research should empirically examine this trade-off to assess the practical viability of stated preference-based algorithms in real-world deployments, particularly in terms of platform incentives and business models.

\subsection{Limitations and Future Work}
We acknowledge several limitations. First, stated preferences measured in this study may not accurately reflect actual preferences of users. Indeed, `true' preferences may be impossible to ascertain. However, we argue that stated preferences, elicited through coin allocation tasks or feed curation activities in which participants are given sufficient time to reflect, serve as a closer proxy than those that are revealed through behaviors. 

A second limitation concerns how to elicit stated preferences without triggering social desirability biases. To mitigate this, the survey avoided explicitly labeling categories like “divisive content” or directly asking how much participants wanted to see them \cite{RathjeEtAl2024a}. While this approach may introduce some noise, it reduces the pressure to conform to socially acceptable responses. Furthermore, in interviews, participants curated feeds for third-party personas, allowing them to articulate what they think more openly.

The remaining limitations relate to ecological and external validity.
The actual social media news feeds of the participants may be very different from the ones we used in this study. Although this choice was made on purpose, as it allowed us to give participants feeds they could interpret easily, these feeds do not capture the full complexity of actual recommendation algorithms. Accordingly, our findings should be interpreted in light of users’ capacity to reason about value trade-offs under the simplified and reflective conditions that our research design enabled, rather than as a direct account of behavior in real-world feed environments or a representation of how feeds operate in practice. 

Related to this point, although this study intentionally employed third-party personas to reduce social desirability bias, we acknowledge that this choice may have further distanced the responses of participants from how they would reason about their own feeds in everyday contexts. Therefore, future studies should investigate how social media users reflect and reason about their actual feeds in more naturalistic settings.

Finally, it is important to note that our study lacks broad external validity outside of the demographic segment under study (young adults). Future research should investigate how the gap between stated and revealed preferences manifests itself among various age demographics of social media users. 

\section{\MakeUppercase{Conclusion}}
This study examined how young adult social media users navigate the gap between stated and revealed preferences, and how algorithms might bridge it to better align with their values. Participants interpreted this gap as a structural mismatch between platform incentives and user values, and found stated preference–based alternatives more satisfying and socially appropriate. In designing ideal news feeds for imagined personas, participants demonstrated a capacity to articulate and reason about what values should be visible and appropriate in shared information spaces. Based on these reflections, participants proposed design ideas for value-aligned recommender systems.

\clearpage
\bibliographystyle{ACM-Reference-Format}
\bibliography{references}

\clearpage

\appendix 

\section*{\MakeUppercase{Appendix}}
\section{\MakeUppercase{Validation of Stated Preference Measurement in Survey}}

One concern regarding the measurement of stated preferences was whether participants interpreted the coin allocation task as intended—that is, whether they allocated coins based on their underlying values rather than incidental factors.

\begin{figure}[ht]
    \centering
    \includegraphics[width=0.8\textwidth]{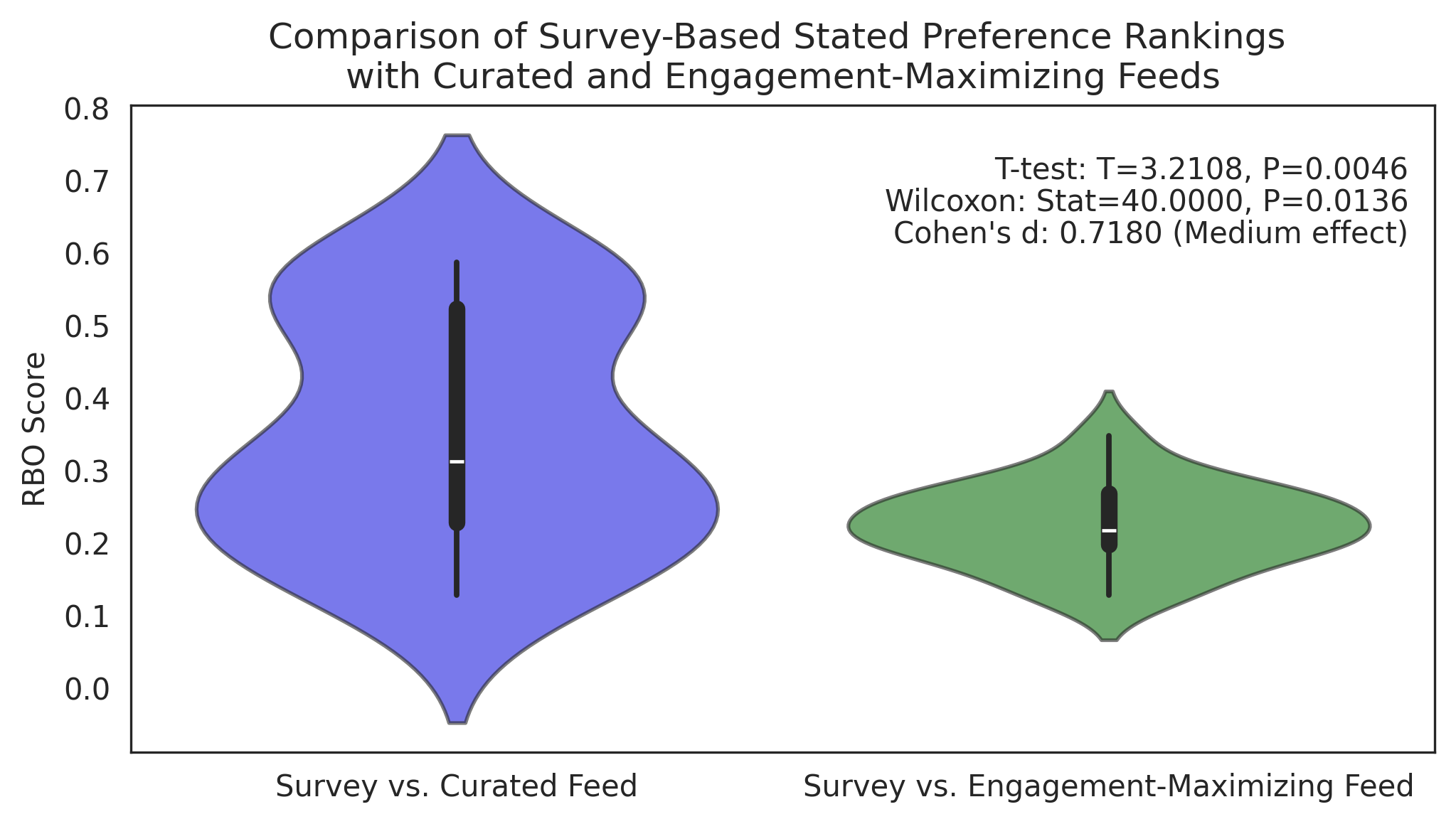}
    \caption{Validating survey-based stated preferences through RBO score comparison}
    \label{fig:Fig9}
    \Description{Figure 9. Validating survey-based stated preferences through RBO score comparison}
\end{figure}

We examined how well participants’ stated preferences from the survey aligned with their curated feeds from the interview. Based on the number of coins allocated in the survey, we constructed a ranked list of content types (i.e., a survey-based feed). We then calculated two RBO scores: one comparing the survey-based feed with the curated feed, and another comparing it with the engagement-maximizing feed. As shown in Fig.~\ref{fig:Fig9}, the RBO score was higher for the survey–curated comparison than for the survey–engagement-maximizing comparison. This suggests that participants’ stated preferences more closely reflect how they curated their feeds based on their underlying values.

For some participants (N = 14; 70\%), we also asked them to reconstruct their thought process during the coin allocation task to check whether their allocation decisions were really driven by stated preferences. 
Participants consistently reported allocating coins based on their underlying values rather than immediate reactions or other biases. For example, P7 prioritized content aligned with “\emph{humanitarian ideals}” and social engagement, while P12 allocated coins to “\emph{science ones or mainstream news}.” 

\section{\MakeUppercase{Supplementary materials}}

\subsection{Survey}\label{appendix:survey}
The full survey questionnaire is available \href{https://drive.google.com/file/d/1QaKk876Kms0M2vZjK-Tule8rO7j3ITSm/view?usp=sharing}{here}.

\subsection{Interview Protocol}\label{appendix:protocol}
\begin{figure} 
  \centering
  \includegraphics[height=0.82\textheight]{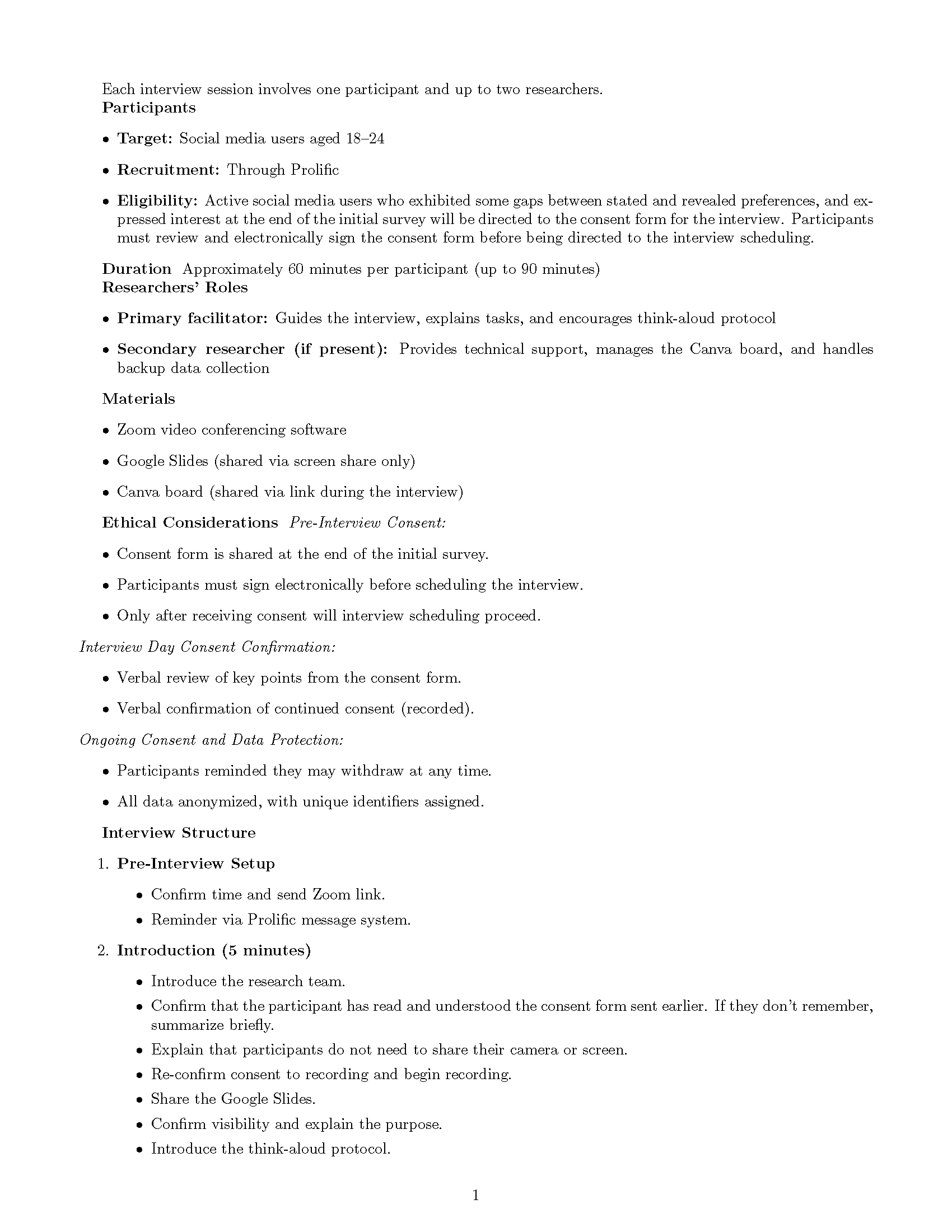}
  \caption{Interview protocol (Page 1)}
  \Description{Interview protocol (Page 1)}
  \label{fig:interview_protocol_1}
\end{figure}

\begin{figure} 
  \centering
  \includegraphics[height=0.82\textheight]{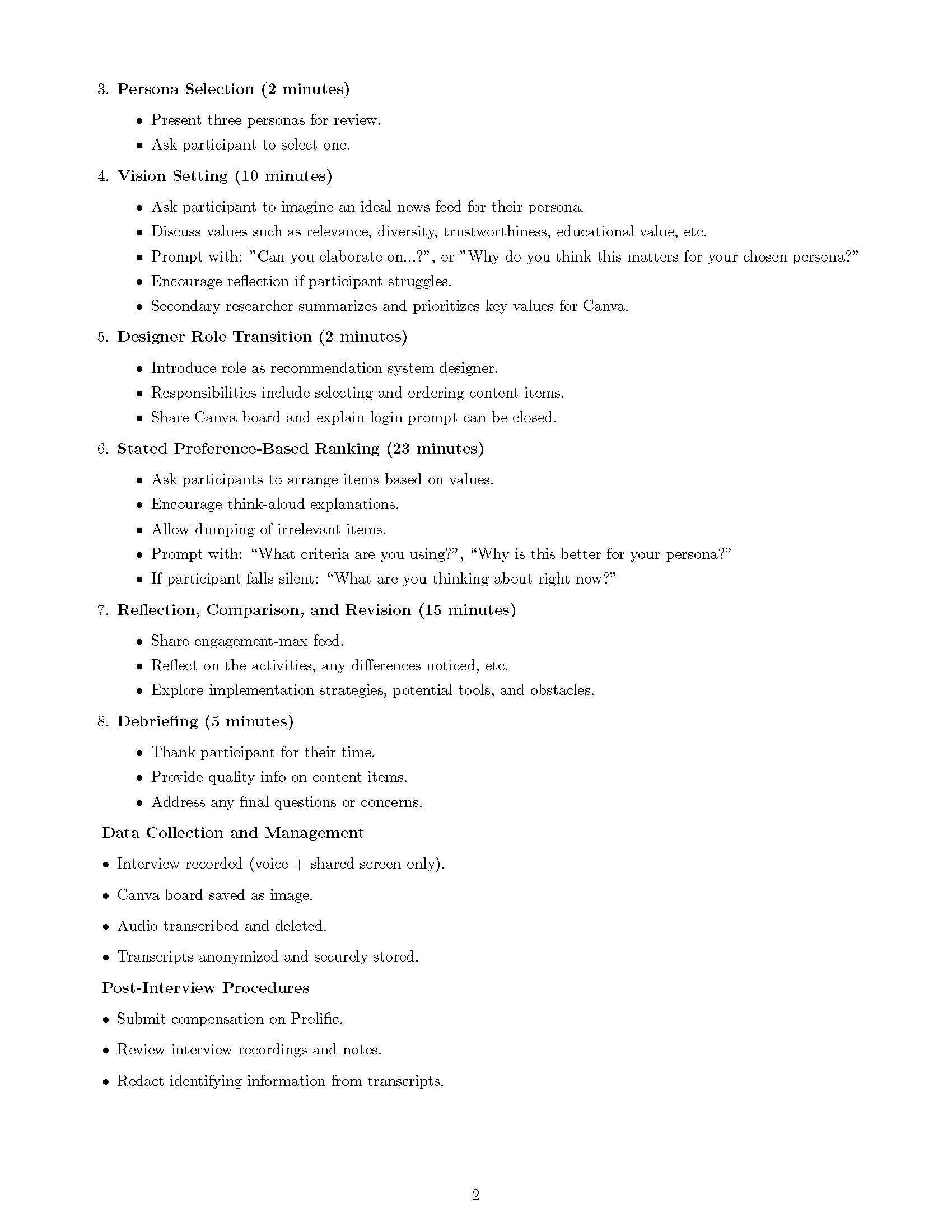}
  \caption{Interview protocol (Page 2)}
  \Description{Interview protocol (Page 2)}
  \label{fig:interview_protocol_2}
\end{figure}

\clearpage

\subsection{Codebook} \label{appendix:codebook}

\begin{table}[ht]
\centering
\resizebox{\columnwidth}{!}{%
\begin{tabular}{|l|l|l|l|}
\hline
\rowcolor[HTML]{EFEFEF} 
\textbf{RQ} &
  \textbf{Themes} &
  \textbf{Sub-Codes} &
  \textbf{Description} \\ \hline
 &
   &
  \begin{tabular}[c]{@{}l@{}}Engagement-max \\ Feed Comparison\end{tabular} &
  \begin{tabular}[c]{@{}l@{}}Perceived differences between participants' \\ curated and engagement-max feeds.\end{tabular} \\ \cline{3-4} 
 &
  \multirow{-3}{*}{\begin{tabular}[c]{@{}l@{}}Making \\ sense\end{tabular}} &
  \begin{tabular}[c]{@{}l@{}}Justification \\ for Gaps\end{tabular} &
  \begin{tabular}[c]{@{}l@{}}Why the Engagement-max Feed \\ diverged from their curated feed.\end{tabular} \\ \cline{2-4} 
 &
   &
  \begin{tabular}[c]{@{}l@{}}Feelings about \\ the Gap\end{tabular} &
  Reflections on the gap \\ \cline{3-4} 
\multirow{-8}{*}{\begin{tabular}[c]{@{}l@{}}RQ2. How do  \\ users make sense \\ of the gap?\end{tabular}} &
  \multirow{-3}{*}{\begin{tabular}[c]{@{}l@{}}Reflecting\\ Discrepancy\end{tabular}} &
  \begin{tabular}[c]{@{}l@{}}Post-Exercise \\ Mental Model\end{tabular} &
  \begin{tabular}[c]{@{}l@{}}Reflections on how their understanding \\ of algorithms evolved\end{tabular} \\ \hline
 &
   &
  Balance, Diversity &
   \\ \cline{3-3}
 &
   &
  Trustworthy, Factual &
   \\ \cline{3-3}
 &
   &
  Scientific, Educational &
   \\ \cline{3-3}
 &
   &
  Sensitivity &
   \\ \cline{3-3}
 &
   &
  Relevance &
   \\ \cline{3-3}
 &
   &
  Entertainment &
   \\ \cline{3-3}
 &
  \multirow{-7}{*}{\begin{tabular}[c]{@{}l@{}}Articulated \\ Values\end{tabular}} &
  Other &
  \multirow{-7}{*}{\begin{tabular}[c]{@{}l@{}}Value dimensions that appeared \\ when participants curated their \\ ideal social media news feed.\end{tabular}} \\ \cline{2-4} 
 &
   &
  Dumped &
   \\ \cline{3-3}
 &
   &
  Ranked High &
   \\ \cline{3-3}
 &
   &
  Ranked Low &
   \\ \cline{3-3}
 &
  \multirow{-4}{*}{\begin{tabular}[c]{@{}l@{}}Curation \\ Decisions\end{tabular}} &
  Signals/Cues &
  \multirow{-4}{*}{\begin{tabular}[c]{@{}l@{}}Decisions participants made during the \\ feed curation process, including why \\ they excluded or prioritized content\end{tabular}} \\ \cline{2-4} 
 &
   &
  Compromises &
  \begin{tabular}[c]{@{}l@{}}Where participants find a middle ground \\ or make conscious trade-offs\end{tabular} \\ \cline{3-4} 
 &
   &
  \begin{tabular}[c]{@{}l@{}}Prioritization \\ of Values\end{tabular} &
  \begin{tabular}[c]{@{}l@{}}Strategies participants use to prioritize \\ certain values over others\end{tabular} \\ \cline{3-4} 
 &
  \multirow{-4}{*}{\begin{tabular}[c]{@{}l@{}}Value \\ Tensions\end{tabular}} &
  \begin{tabular}[c]{@{}l@{}}Trade-offs Between \\ Values\end{tabular} &
  \begin{tabular}[c]{@{}l@{}}Conflicts participants face when \\ trying to balance competing values\end{tabular} \\ \cline{2-4} 
 &
   &
  \begin{tabular}[c]{@{}l@{}}Network Tags,\\ Social Ties\end{tabular} &
  Influence of social ties on curation \\ \cline{3-4} 
 &
   &
  Partisan predisposition &
  Influence of users' predisposition \\ \cline{3-4} 
 &
   &
  Persona Influence &
  Influence of the persona's characteristics \\ \cline{3-4} 
\multirow{-25}{*}{\begin{tabular}[c]{@{}l@{}}RQ3. What value \\ dimensions do \\ users consider, \\ and how do they \\ navigate tensions \\ between values?\end{tabular}} &
  \multirow{-4}{*}{\begin{tabular}[c]{@{}l@{}}Decision \\ Context\end{tabular}} &
  \begin{tabular}[c]{@{}l@{}}Pre-existing knowledge \\ about sources\end{tabular} &
  \begin{tabular}[c]{@{}l@{}}Influence of user's existing knowledge \\ about different sources\end{tabular} \\ \hline
 &
   &
  \begin{tabular}[c]{@{}l@{}}Adjustable vs. \\ Fixed Algorithms\end{tabular} &
  \begin{tabular}[c]{@{}l@{}}Allowing users to switch between \\ different algorithmic modes\end{tabular} \\ \cline{3-4} 
 &
   &
  \begin{tabular}[c]{@{}l@{}}Customization \\ Options\end{tabular} &
  \begin{tabular}[c]{@{}l@{}}Tools that allow users to fine-tune \\ the system according to their preferences\end{tabular} \\ \cline{3-4} 
 &
   &
  Stated Preference &
  Ways to incorporate stated preferences \\ \cline{3-4} 
 &
   &
  Transparency &
  \begin{tabular}[c]{@{}l@{}}Clarity in how the algorithm works, \\ explanations on why certain content appears\end{tabular} \\ \cline{3-4} 
 &
   &
  User Control &
  \begin{tabular}[c]{@{}l@{}}Features that allow users to manually \\ curate or adjust their feed\end{tabular} \\ \cline{3-4} 
 &
  \multirow{-10}{*}{\begin{tabular}[c]{@{}l@{}}Articulated\\ Values\end{tabular}} &
  Other &
  Ideas related to media literacy skills \\ \cline{2-4} 
 &
   &
  Challenges &
  \begin{tabular}[c]{@{}l@{}}Obstacles or barriers participants identify in \\ implementing their proposed ideas\end{tabular} \\ \cline{3-4} 
\multirow{-12}{*}{\begin{tabular}[c]{@{}l@{}}RQ4. How can \\ recommendation \\ systems be \\ designed to better \\ align with \\ stated preferences?\end{tabular}} &
  \multirow{-3}{*}{\begin{tabular}[c]{@{}l@{}}Challenges, \\ Opportunities\end{tabular}} &
  Opportunities &
  Potential innovations and ideas suggested \\ \hline
\end{tabular}%
}
\end{table}

\end{document}